\documentclass[apj]{emulateapj}

\usepackage{amsmath}
\usepackage{xspace} 
\usepackage{graphicx}
\usepackage{txfonts}
\usepackage{rotate}
\usepackage{fancyheadings}

\newcommand{\un}[1]{~\hspace{-0.8pt}\ensuremath{\mathrm{#1}}}
\long\def\symbolfootnote[#1]#2{\begingroup\def\thefootnote{\fnsymbol{footnote}}\footnote[#1]{#2}\endgroup}

\newcommand{\integ}{{\it Integral}\xspace}
\newcommand{\ibis}{Ibis\xspace}
\newcommand{\isgri}{Isgri\xspace}
\newcommand{\spi}{Spi\xspace}
\newcommand{\jemx}{Jem-X\xspace}
\newcommand{\xmm}{{\it XMM-Newton}\xspace}
\newcommand{\epic}{Epic\xspace}
\newcommand{\mos}{Mos\xspace}
\newcommand{\pn}{{\sc\large pn\/}\xspace}
\newcommand{\chandra}{{\it Chandra}\xspace}
\newcommand{\asca}{{\it Asca}\xspace}
\newcommand{\hess}{{\it Hess}\xspace}
\newcommand{\granat}{{\it Granat}\xspace}
\newcommand{\artp}{Art-P\xspace}
\newcommand{\sigm}{Sigma\xspace}
\newcommand{\xrt}{{\sc\large xrt\/}\xspace}
\newcommand{\egret}{{\it Egret}\xspace}
\newcommand{\cgro}{{\sc\large cgro\/}\xspace}
\newcommand{\vla}{{\sc\large vla\/}\xspace}
\newcommand{\spacelab}{{\it Spacelab\,2}\xspace}

\newcommand{\sgra}{Sgr\,A$^{*}$\xspace}
\newcommand{\sgraeast}{Sgr\,A\;East\xspace}
\newcommand{\igr}{IGR\,J17456--2901\xspace}

\newcommand{\cxo}{CXOGC\,J174540.0--290031\xspace}
\newcommand{\cxogc}{CXOGC\,J}

\newcommand{\gammarays}{$\gamma$-rays\xspace}
\newcommand{\gammaray}{$\gamma$-ray\xspace}
\newcommand{\xray}{X-ray\xspace}
\newcommand{\xrays}{X-rays\xspace}
\newcommand{\fwhm}{{\sc\large fwhm\/}\xspace}
\newcommand{\gti}{{\sc\large gti\/}\xspace}
\newcommand{\psla}{{\sc\large psla\/}\xspace}
\newcommand{\pwn}{{\sc\large pwn\/}\xspace}
\newcommand{\snr}{{\sc\large snr\/}\xspace}
\newcommand{\gc}{{\sc\large gc\/}\xspace}
\newcommand{\gn}{{\sc\large gc\/}\xspace}
\newcommand{\psf}{{\sc\large psf\/}\xspace}

\newcommand{\go}{{\sc\large go\/}\xspace}
\newcommand{\xspec}{Xspec\xspace}
\newcommand{\ir}{{\sc\large ir}\xspace}
\newcommand{\nir}{{\sc\large nir}\xspace}
\newcommand{\uv}{{\sc\large uv}\xspace}
\newcommand{\bh}{{\sc\large bh}\xspace}
\newcommand{\osa}{{\sc\large osa}\xspace}
\newcommand{\isdc}{{\sc\large isdc\/}\xspace}
\newcommand{\lmxb}{{\sc\large lmxb\/}\xspace}

\newcommand{\am}{$^{\prime}$\xspace}
\newcommand{\as}{$^{\prime\prime}$\xspace}
\newcommand{\ergpersec}{\;erg\,s$^{-1}$\xspace}
\newcommand{\ergpercmsec}{\;erg\,cm$^{-2}$\,s$^{-1}$\xspace}
\newcommand{\phcmskev}{\;ph\,cm$^{-2}$\,s$^{-1}$\,keV$^{-1}$\xspace}
\newcommand{\phpercmpersec}{\;ph\,cm$^{-2}$\,s$^{-1}$\xspace}
\newcommand{\degree}{$^{\circ}$\xspace}

\newcommand{\bk}{\hspace{-5pt}}
\newcommand{\aplc}{ApLC\xspace}

\submitted{2005 May 28}
\received{2005 May 31}
\accepted{2005 July 21}

\begin{document}

\title{A Persistent High-energy Flux from  the Heart of the Milky Way:\\ Integral's view of the Galactic Center\footnotemark[*]} \footnotetext[*]{Based on observations with INTEGRAL, an ESA project with instruments and science data center funded by ESA member states (especially the PI countries: Denmark, France, Germany, Italy, Switzerland, Spain), Czech Republic and Poland, and with the participation of Russia and the USA.}

\author{
	G.\ B\'elanger\altaffilmark{1,2}, A.\ Goldwurm\altaffilmark{1,2}, 
	M.\ Renaud\altaffilmark{1,2}, 
	R.\ Terrier\altaffilmark{1,2}, 
	F.\ Melia\altaffilmark{3},
	N.\ Lund\altaffilmark{4}, 
	J.\ Paul\altaffilmark{1,2}, 
	G.\ Skinner\altaffilmark{5},  
	F.\ Yusef-Zadeh\altaffilmark{6}
	}

\altaffiltext{1}{\scriptsize Service d'Astrophysique, DAPNIA/DSM/CEA, 
			91191 Gif-sur-Yvette, France; belanger@cea.fr}
\altaffiltext{2}{\scriptsize Unit\'e mixte de recherche Astroparticule et Cosmologie, 
			11 place Berthelot, 75005 Paris, France}
\altaffiltext{3}{\scriptsize Physics Dept. and Stewart Observatory, Univerity of Arizona, 
			Tucson, AZ 85721, USA; melia@physics.arizona.edu}
\altaffiltext{4}{\scriptsize Danish National Space Center, Juliane Maries vej 30,
			Copenhagen, Denmark; nl@spacecenter.dk}
\altaffiltext{5}{\scriptsize CESR, Toulouse Cedex\,4, France; skinner@.cesr.fr}
\altaffiltext{6}{\scriptsize Department of Physics and Astronomy, Northwestern University, 
			Evanston, IL 60208}

\begin{abstract}
Highly sensitive imaging observations of the Galactic center (\gc) at high energies 
with an angular resolution of order 10 arcminutes, is a very recent development 
in the field of high-energy astrophysics. The \ibis/\isgri imager on the \integ observatory 
detected for the first time a hard \xray source, \igr, located within 1 arcminute of 
Sagittarius\,A$^{*}$ (\sgra) over the energy range 20--100\un{keV}.
Here we present the results of a detailed analysis of approximately 7\,$\times$\,$10^{6}$\un{s} 
of observations of the \gc obtained since the launch of \integ in October 2002.
Two years and an effective exposure of 4.7\,$\times$\,$10^{6}$\un{s}
have allowed us to obtain more stringent positional constraints on this high-energy source
and to construct its spectrum in the range 20--400\un{keV}.
Furthermore, by combining the \isgri spectrum together with the total 
\xray spectrum corresponding to the same physical region around \sgra from \xmm data, 
and collected during part of the gamma-ray observations, we constructed and 
present the first accurate wide band high-energy spectrum for the central 
arcminutes of the Galaxy. Our complete and updated analysis of the emission properties of 
the \integ source shows that it is faint but persistent with no variability above 3\,$\sigma$, 
contrary to what was alluded to in our first paper.  
This result, together with the spectral characteristics of the soft and hard \xray emission from 
this region, suggests that the source is most likely not point-like but, rather, 
that it is a compact, yet diffuse, non-thermal emission region.
The centroid of \igr is estimated to be R.A.\,=\,$\mathrm{17^{h}45^{m}42^{s}\hspace{-2pt}.5}$, 
decl.\,=\,$-28^{\circ}59'28''$ (J2000),
offset by 1\am from the radio position of \sgra and with a positional uncertainty of 1\am. 
Its 20--400\un{keV} luminosity at 8\un{kpc} is
$L$\,=\,(5.37\,$\pm$\,0.21)\,$\times$\,$10^{35}$\ergpersec. 
A 3\,$\sigma$ upper limit on the flux at the electron-positron annihilation energy of 511\un{keV} 
from the direction of \sgra is set at 1.9\,$\times$\,$10^{-4}$\phpercmpersec. 
Very recently, the \hess collaboration presented the detection of a source of $\sim$\,TeV 
\gammarays also located within an arcminute of \sgra. 
We present arguments in favor of an 
interpretation according to which the photons detected by \integ and \hess arise from 
the same compact region of diffuse emission near the central black hole 
and that the supernova remnant \sgraeast could play an important role as a 
contributor of very high-energy \gammarays to the overall spectrum from this region.
There is also evidence for hard emission from a region located between the central black
hole and the radio Arc near l\,$\sim$\,0.1\degree along the Galactic plane and known
to contain giant molecular clouds.
\end{abstract}


\section{Introduction}
\label{s:intro}
The year 2004 marked the $30^{\rm th}$ anniversary of the discovery of the compact 
radio source \sgra\,\citep{c:balick74}, which is now firmly believed by many to be the 
manifestation of a supermassive black hole that sits at the very heart 
of the Milky Way and around which everything in the Galaxy turns.
That year also marked the first detection of \gammarays from a compact region 
of size $\sim$\,10 arcminutes around \sgra with the \integ observatory in the energy 
range from 20 to 100\un{keV} \citep{c:belanger04} and with the \hess Cerenkov telescope
array between 165\un{GeV} and 10\un{TeV} \citep{c:aharonian04}.  After three decades of 
observations, we finally detected a source of very high-energy radiation that 
appears to be point-like and coincident with the Galactic nucleus (\gn).  However, the exact 
nature of the highly energetic emission from this compact region is unknown.  Our aim 
here is to present observational evidence that will lead to a deeper understanding of 
the emission process, and help to unfold the mystery of the \gammarays arriving from
the heart of the Milky Way.

The Galactic nuclear region is very dense and complex --- so dense that certainly more
than one source could contribute to the high-energy flux detected by present-day 
\gammaray instruments with typical angular resolutions of 10--15 arcminutes.
Located at 8\un{kpc} from the Sun, the Galactic center (\gc) harbours a supermassive 
black hole whose presence and mass of about 3.6\,$\times$\,$10^{6}$\un{M_{\odot}} 
were deduced primarily from near-infrared (\nir) observations and measurements of the 
velocity and proper motion of the stars contained in the central cluster 
\citep{c:schodel03, c:ghez05, c:eisenhauer05}.
A black hole (\bh) of this mass has a Schwarzschild radius ($R_\mathrm{S}$) of about 
1.2\,$\times$\,$10^{12}$\un{cm} and is expected to accrete the matter from its nearby 
environment producing a detectable emission in a broad range of frequencies \citep{c:melia-falcke01}.
The bright, compact, non-thermal radio source \sgra,
located at less than 0\as\bk.01 from the dynamical center of the
central star cluster, is most likely the manifestation of such accretion processes.
Undetectable in the visible and \uv bands due to the large Galactic absorption,
and only recently detected in \nir both in quiescent, 
and flaring states \citep{c:genzel03, c:ghez04}, this source is surprisingly weak in \xrays,
where it appears slightly extended with a luminosity of only
$L_{\rm X}${\footnotesize[2--10\un{keV}]}\;$\approx$\;2\,$\times$\,$10^{33}$\ergpersec \citep{c:baganoff03}.

The dense central region of the Milky Way where the 2--10\un{keV} flux
is heavily dominated by diffuse radiation, 
includes the emission from a hot ($kT$\,$\sim$\,6--8\un{keV}) probably unbound 
thermal plasma component, the supernova remnant (\snr) \sgraeast, 
several knots and filaments, and thousands of point sources
\citep{c:maeda02, c:park04, c:muno04a, c:muno04b}. 
The faint \xray counterpart of \sgra would likely have gone unnoticed were it not
positioned almost exactly at the centre of the Galaxy.

Following a monitoring of this faint \xray source by the \chandra observatory,
it was discovered that \sgra is the site of sometimes powerful \xray flares 
\citep{c:baganoff01}. This flaring activity was also detected and studied 
with \xmm \citep{c:goldwurm03a,c:porquet03a,c:belanger05} and the peak luminosity
during these flares was seen to rise above quiescence by factors up to 180
and then decay in a few hours or less. The majority of flares have spectra that 
are significantly harder (power law photon index $\Gamma$\,$\sim$\,1.5) 
than the quiescent spectrum ($\Gamma$\,$\sim$\,2.7),
but the most powerful of them was quite soft ($\Gamma$\,$\sim$\,2.5).
Note that dust scattering can have some effect on the observed fluxes and 
spectral indeces of both the quiescent and flaring states of \sgra \citep{c:tan-draine04}.
Variations on time scales as short as 200\un{s} have been detected during flares 
indicating an emitting region with a size of the order of 10$R_\mathrm{S}$. 
Recent \chandra and \xmm campaigns have allowed us to estimate the average \xray 
flare rate to about 1 per day bearing in mind that the flaring events appear to be clustered 
\citep{c:belanger05}. At slightly longer wavelengths, \nir observations with the 
{\it \small VLT\/} Naco imager \citep{c:genzel03} and {\it Keck} telescope \citep{c:ghez04} 
have shown that \sgra is also the source of frequent \ir flares with durations comensurable
with those seen in \xrays. Some \ir flares even appear to be simultaneous with \xray flares 
\citep{c:eckart04}.  Both the \ir and \xray flaring events strongly suggest the
presence of an important population of non-thermal relativistic electrons in the
vicinity of the \bh horizon \citep{c:markoff01, c:liu-melia02, c:yuan02, c:yuan03, c:liu04} 
and therefore their detection has raised great interest in the possibility
of observing hard \xrays from the Galactic nucleus.

\sgra's bolometric luminosity from radio to \xrays (including the flares) 
barely amounts to a few $10^{36}$\ergpersec,
while the Eddington luminosity for a \bh of its mass reaches
$L_{\rm E}$\;$\approx$\;4\,$\times$\,$10^{44}$\ergpersec.
Even the expected accretion luminosity, based on an estimated
stellar wind mass rate at the accretion radius, is of the order of  $10^{42-43}$\ergpersec,
i.e. about 6 to 7 orders of magnitude higher than \sgra's total observed emitted power
(see Cuadra et al.\ 2005 for a recent update on the issue).

In the 1990's it was thought that the bulk of the power could be found
at higher energies; in hard \xrays as is the case for \bh binaries in the hard state, 
or even at the electron-positron annihilation energy of 511\un{keV}
(see e.g. Genzel \& Townes\ 1987).
Other than the detections in the 10--20\un{keV} range from the direction of the \gc
based on \xrt/\spacelab \citep{c:skinner87} and \artp/\granat \citep{c:pavlinsky94} data, 
no detection at energies above 20\un{keV} emanating from Sgr\,A complex was reported.
A long monitoring of the region by the \sigm telescope on the \granat satellite yielded 
upper limits of the order of several $10^{35}$\ergpersec to the 35--150\un{keV} 
emission from \sgra \citep{c:goldwurm94, c:goldoni99}, 
and 2.3\,$\times$\,$10^{-4}$\phpercmpersec to the flux at 511\un{keV} from a point source
at the \gc \citep{c:malet95}.

In the 100\un{MeV} to 10\un{GeV} energy range, an unidentified \gammaray source of the 
\egret/\cgro catalog 3EG\,J1746--2851, 
was found to be somewhat compatible with the \gn \citep{c:mayer98}.
In the third \egret catalog \citep{c:hartman99},
this source is located 0.17\degree from \sgra and the reported
error radius is 0.13\degree at 90\% confidence level.
Taking this at face value would marginally exclude \sgra. However, given
the 1\degree angular resolution of the instrument and hence its inability to exclude
the contribution from other sources contained in a region of this size,
3EG\,J1746--2851 is still considered as possibly coincident with the \gn.

Finally, \integ observations performed during the first half of 2003 with the \ibis telescope 
revealed for the first time the presence of a significant excess in the energy range 
20--100\un{keV} coming from the inner region of the Galaxy \citep{c:belanger04}.
The position of this excess was found to lie 1\am from \sgra with a 4\am error radius,
and its luminosity was estimated to be 
$L_{\rm X}${\footnotesize[20--100\un{keV}]}\;$\approx$\;3\,$\times$\,$10^{35}$\ergpersec.
An indication of variability on timescales comparable to those of flares in \sgra was reported but
a subsequent analysis that included data collected over the second half of 2003 with improved
analysis procedures found the source to be stable \citep{c:goldwurm04}.
In 2003 June, \chandra detected two new transient sources in the close vicinity of \sgra.
Given the 12\am resolution of the \ibis/\integ telescope,
the possibility that the 20--100\un{keV} excess was linked to the emission from these objects,
was considered and discussed in B\'elanger et al.\ (2004).

Meanwhile the \hess collaboration \citep{c:aharonian04} announced the detection of a bright
source of TeV photons coincident with \sgra to within 1\am. The source appears to be point-like, 
stable, with a power-law spectrum of index of 2.2 and a luminosity 
of $L_{\gamma}${\footnotesize[1--10\un{TeV}]}\;$\approx$\;$10^{35}$\ergpersec.

This wealth of high-energy data all point to the presence of one, or several, high energy
non-thermal emission components likely produced by accelerated particles in the
environment of the \gc.
Both leptonic and hadronic origins for the accelerated particles giving rise to the \gammarays
have been considered either in the inner \citep{c:markoff97, c:aharonian-neronov05}
or outer \citep{c:atoyan-dermer04} region of \sgra, 
while Melia et al.\ (1998), Fatuzzo \& Melia (2003) and recently Crocker et al.\ (2005),
discussed the possibility that the site of particle acceleration could be the unusual
supernova remnant \sgraeast.

\sgraeast is a mixed morphology \snr whose center is at less than 1\am from \sgra and
whose radio shell spans a few arcminutes. Given its position and dimensions it would appear 
like a stable point-like source for \gammaray observatories.
This remnant is characterised by a non-thermal radio shell at the center of which lies an 
apparently ejecta-dominated \xray emiting region whose spectrum indicates that the plasma
is rich in heavy elements especially towards the core where Fe abundances reach 4--5 times
solar \citep{c:maeda02, c:sakano04}. 
For this reason, it has been classified as a metal-rich mixed morphology \snr.
The fact that \sgraeast is the smallest of the known \snr\/s of this type
and that its radio shell appears to be quite symmetrical, although slightly elongated along 
the Galactic plane, suggests that the ejecta from the explosion have expanded in a very 
dense but more or less homogenous environment \citep{c:maeda02}. 
According to a more recent analysis of the \xray features of \sgraeast by Sakano et al.\ (2004)
based on \xmm observations, the derived total energy, mass and abundance pattern are consistent 
with a single supernova event of Type Ia or Type II involving a relatively low-mass progenitor 
star. Furthermore, according to these authors the morphology and spectral characteristics do not
show evidence of a clear connection between the \snr and past activity in \sgra.

\sgraeast may very well have arisen from a single explosion akin to most other supernova 
events in terms of its energetics as is suggested by Sakano et al.\ (2004). The most recent
simulations for its genesis and evolution (Fryer et al. 2005) indicate that the progenitor
was likely a star of mass $\sim$\,15\,$M_\odot$ that exploded a mere $\sim$\,1750 years ago.
Reaching the M-0.02-0.07 cloud some 300--400 years ago, the expanding shock
collided with the dense molecular gas, producing a bright flash of 2--200\un{keV} emission 
lasting several hundred years, whose X-ray echo we may be viewing today in the form of
Fe fluorescent emission from Sgr B2 (and other nearby clouds). However, Sgr A East
does distinguish itself from other Galactic remnants in three important ways. 
First, it is located very near the \gc, within 50\as of \sgra, and is therefore subject to 
interactions and forces uncommon in the rest of the Galaxy. 
Second, its non-thermal shell emission caused by synchrotron radiation from relativistic
electrons has an unusually high surface brightness compared to other Galactic \snr \citep{c:green04}. 
Third, OH (1720 MHz) maser emission detected in several locations
around the \gc, and particularly at the boundary of \sgraeast and the M-0.02-0.07 molecular cloud, 
indicates the presence of strong shocks where rapid acceleration of electrons (and protons)
is taking place, in a medium threaded by very strong magnetic fields of order 
2--4\un{mG} \citep{c:yusefZadeh96}.  As pointed out by Yusef-Zadeh, Melia \& Wardle (2000), 
the presence of relativistic electrons and strong magnetic fields within \sgraeast makes it 
a unique and potentially powerful Galactic accelerator.

Other possible sources of non-thermal \gammarays, some of which were initially proposed to 
explain certain features of the \xray emission like the hot component and the
bright 6.4\un{keV} line of neutral Fe, or the non-thermal radio characteristics of the
region, include the radio Arc,
several non-thermal filaments and regions marked by cosmic ray electron
interactions \citep{c:yusefZadeh02}, supernova ejecta \citep{c:bykov02}, 
and scattering of highly energetic radiations from molecular clouds. 
In particular Revnivtsev et al.\ (2004) proposed a model in which the \integ source 
IGR\,J17475--2822, coincident with the dense molecur cloud Sgr\,B2, is due to
Compton reflection (i.e. the scattering of high-energy photons by cold 
electrons in the outer layers of the cloud) of very energetic emission from a 
very powerful flare in the \sgra system about 300 years ago.
In any case, none of the current models can integrate the three high-energy 
sources  (\igr, 3EG J1746-2851, HESS\;J1745--290) detected near the \gc in a
comprehensive manner.

We report here a complete study of the 2003-2004 \integ/\ibis data of the \gc 
aiming to clearly depict the morphology of this interesting region at energies
20 to 400\un{keV}, and to present the properties of the central source \igr. 
The \integ observatory monitored the \gc region for all of 2003 and 2004, including
some dedicated programs specifically planned to study the properties of the \gn.
An important goal of this study was to search for correlated variability between the
\xrays from \sgra and the higher energy emission from the central \integ source.

We describe the observations and data reduction methods in sections \ref{s:observations} 
and \ref{s:analysis}. 
Our results are presented in \textsection\,\ref{s:results} and include those of 
the multi-wavelength campaign on \sgra performed in 2004 pertaining to \integ 
(\textsection\,\ref{s:multiwavelength}). 
Finally, we discuss some of their implications in \textsection\,\ref{s:discussion}.

\section{Observations}
\label{s:observations}
The \integ observatory carries two main \gammaray instruments; \ibis and \spi,
working in the energy range 15\un{keV} to 10\un{MeV} \citep{c:winkler03}.
Since its launch, \integ has observed the central degrees of our Galaxy
for a total of about 7\un{Ms}. This time was divided  between the Galactic 
Center Deep Exposure and Galactic Plane Scan core programs, 
and Guest Observer (\go) observations.  
In particular we include here analysis of \go programs performed in 
2003 and in 2004 specifically dedicated to the \gn.
The 2004 \go program ($\sim$\,600 ks) was part of a broad multiwavelength 
campaign driven by a \xmm large project aimed to study the 
flaring activity of \sgra.

The data that form the basis of this 
paper constitute a subset of all these observations 
selected such that the aim point is within 10\degree of the central \bh.
We have performed a detailed analysis of 2174 pointings carried out between 
the end of February 2003 and October 2004. Each pointing or science window 
(ScW) typically lasts between 1800--3600\un{s} during which the telescopes are aimed
at a fixed direction in the sky. Table\,\ref{t:obsLog} gives a summary of the 
overall periods spanned by the observations.

With a total effective exposure time of 4.7\un{Ms} at the position of \sgra,
we have constructed high signal-to-noise images of the 
central degrees and the spectrum of the central source \igr first detected by \integ in 
2003 \citep{c:belanger04}. The results presented here are based on data 
collected with the \ibis/\isgri telescope \citep{c:ubertini03, c:lebrun03}
sensitive in the energy range between 15 and 1000\un{keV}. 
The angular resolution of the high-energy (15\un{keV}--10\un{MeV})
\spi telescope ($\sim$3\degree) is not sufficient to resolve the contribution of the 
high-energy sources known to be present in the central degrees of the Galaxy and we 
have therefore not used these data. 

The \xray monitor on-board \integ, \jemx, has a smaller field of view
than \ibis and \spi and therefore the effective exposure at the location of \sgra
is substantially lower ($\sim$\,350\un{ks} for the  dataset considered in this paper). 
We analyzed the \jemx data to produce mosaics for the \gc and discuss the results 
below.

We have also made use of \xmm data from the multi-wavelength campaign on \sgra 
carried out in 2004 to construct the broadband high-energy spectrum of \igr, and to 
identify the \xray and soft \gammaray components.  A detailed description of the 
\xmm observations during this campaign, and of the characteristics of the two 
factor-40 flares detected from the direction of \sgra are presented in B\'elanger et al.\ (2005).

\begin{deluxetable}{lcccc}
\tablewidth{0pt}
\tablecolumns{4}
\tablecaption{Observation Log}
\tablehead{
& 	\colhead{Obs Start time}	& \colhead{Obs End time}	& \colhead{Pointings}	& \colhead{Exposure} \\
  \colhead{} &    \colhead{(UT)}	&    \colhead{(UT)}		& 			& \colhead{(Ms)}
}
\startdata
2003 Spring$^{\un{a}}$	& 2003-02-28	&  2003-04-22 		&	413		& 0.67 \\
2003 Fall$^{\un{b}}$	& 2003-08-18	&  2003-10-16 		&	805		& 2.20 \\
2004 Spring		& 2004-02-07	&  2004-04-21 		&	550		& 1.01 \\
2004 Fall			& 2004-08-28	&  2004-09-17 		&	262		& 0.53
\enddata
\tablenotetext{a}{Loosely used to designate the first part of the year}
\tablenotetext{b}{Used to designate the second part of the year}

\label{t:obsLog}
\end{deluxetable}

\section{Data Analysis Methods}
\label{s:analysis}
The basic \ibis/\isgri data reduction for individual pointings was performed using the 
\integ Off-line Scientific Analysis software (\osa) version 4.2
delivered by the \isdc \citep{c:courvoisier03} and whose algorithms are described 
in Goldwurm et al.\ (2003b) and Gros et al.\ (2003).
Following the recommendations related to the use of the \osa software,
we restricted our analysis to events with energies greater than 20\un{keV}.
A number of additional procedures were implemented in order to maximize the quality
of our analysis given the large data set and the complexity of Galactic nuclear region. 
For instance, the analysis was consistently done twice: a first time to make a catalog 
of detected sources and perform a preliminary evaluation of the quality of each sky 
image, and a second time to ensure that all known sources in a given field of view were 
modeled correctly in the reconstruction of each sky image.
The total number of sources detected by \isgri within 10\degree of the \gc, and therefore
included in the analysis input catalog is 80.
The background maps were constructed from empty field observations at high latitudes
in 256 bands from 17 to 1000\un{keV} and incorporated in the standard analysis
where they were combined to match the chosen energy ranges.
This procedure has allowed us to achieve the best possible 
quality for individual sky images and thus for the mosaic and spectrum.
Other additional analysis procedures are detailed in the sections that follow.

\subsection{Sky maps, light curves and mosaics}
For a coded mask instrument like \ibis, a sky image is obtained by convolving the
detector image or shadowgram with a decoding array derived from the spatial 
characteristics of the mask.
For each sky image there is an intensity map ($I$), a variance map ($V$) 
proportional to the total counts recorded in each pixel, 
and a significance map ($S$) constructed from these as $S=I/\sqrt{V}$.
Given that source photons are a very small fraction of the total counts, 
the background heavily dominates and the histogram of significance values for 
a given deconvolved image should follow the standard Normal distribution.
Significant detections appear as spikes in the positive tail of the distribution.
A broadening of this distribution is caused by systematic 
effects unaccounted for in the standard analysis software.
Therefore, the standard deviation of the distribution of significance values
can be effectively used to characterize and quantify the quality
of a given sky image, and must be taken into account when calculating the true
detection significance of a signal in that image\,\footnote[$\dagger$]
{ 
The method used to account for systematic effects
on a per-ScW basis described here is equivalent to deriving
a correction factor to the theoretical statistical uncertainty and applying it
to each pixel in the variance map. It follows that any test statistic such as
Pearson's $\chi^{2}$-test in which the variance of each data point is used to
weigh the associated data value when testing for a deviation from the mean, 
should be used with the corrected variance values, 
i.e. the statistical variance multiplied by the variance of the distribution
of significance values.
}.
For this reason, we have weighted each sky image according to the variance of 
its significance distribution to produce sky maps in seven energy bands from 20 to 400 keV; 
namely 20--30, 30--40, 40--56, 56--85, 85--120, 120--200, 200--400\un{keV},
and to construct the spectrum of the \gc source \igr. 
We also performed the analysis of another energy band from 500 to 522\un{keV}
which yielded upper limits compatible with the \spi detection of the 
electron-positron 511\un{keV} annihilation line reported in Knodlseder et al.\ (2005)
(see \textsection\,\ref{s:results}).
The mosaics were made using the {\it pixel spread} option in the \osa\,4.2
imaging procedure which by projecting and distributing pixel values on the
final mosaic pixel grid, preserves the symmetry of the point spread function (\psf) 
and allows the most accurate source positioning.

Since the central source is too weak to be fitted with the \psf in the imaging
procedure performed on each pointing, the light curve was obtained by extracting 
the count rate at the source's pixel position from each reconstructed image.

\subsection{Source position determination}
\label{s:position}
The densely populated region around \sgra contains at least eight sources 
detected by \isgri within 1 degree of the \gn. Since the standard \osa pipeline
does not perform simultaneous fitting of several sources, we have used a 
custom fitting procedure to determine the best fit positions of the sources
in the neighborhood of the \gc. Sub-images 40\,$\times$\,40 pixels in size centered
on the radio position of \sgra were extracted from the mosaics in the different
energy bands and fitted using a model that included up to eight point-sources, 
each characterised by a 2\-D Gaussian approximation of the system point spread
function and applied in the \osa software \citep{c:gros03}

The width of the \psf was left as free global parameter (1 for all sources) 
because for mosaics, the width of the final \psf cannot be predicted. 
The presence of close sources and possible confusion does not influence the result
since a very strong point-like source (1E\,1740.7--2942) dominates the 
images and this parameter is basically determined by the fit of this source.
The procedure gives us the possibility to fix the position of some sources
and also to set a flat background level to be fitted together with the point-sources.
The residual map is inspected to verify that the fitting procedure 
is performed correctly. 

Statistical errors at 90\% confidence level on the fitted positions were derived 
from the measured source signal-to-noise ratios using the empirical law determined 
through a systematic study of well known sources \citep{c:gros03}.
Although the empirical function of the source location accuracy in terms of the
detection significance was determined using images reconstructed from single 
pointings, it was recently validated on mosaics as well. These tests were 
performed on isolated point-sources and since we are studying a region where the 
source density is unusually high, we present the results of our simultaneous fit for 
all the sources in the field to show that the offsets from the known positions are 
compatible with the empirical point-source location accuracy (\psla).

\subsection{Spectral extraction}
\label{s:spectralExtraction}
The standard \isgri spectral extraction software works on the basis of
a single pointing in the following manner. Using a list of sources, 
the procedure first calculates the pixel illumination factor;
a model of the shadow of the mask cast on the detector plane by a given source.
With this set of modeled shadowgrams, the software then attempts to
fit the detector image by adjusting the relative intensity
of each source either together with the background level simultaneously
or by subtracting a background map normalized to the mean count rate beforehand.
These procedures therefore estimate the maximum likelihood intensity 
of each source and in each pointing.
However, a difficulty arises in the case of week sources ($\sim$mCrab) 
that require very long exposure times to acquire sufficient statistics to be 
detected. In such cases, and in particular when a large number of sources have to 
be modelled, the most efficient way to derive a spectrum is to build and 
clean sky images in the desired energy bands and then to extract in each 
the intensity and its associated variance at the pixel that corresponds 
to the source position, and to calculate the weighted mean count rate.
As was described in the previous section, the variance at a given 
pixel must be corrected to take into account the systematic effects that grow 
additively with exposure time. In the case of \igr, 
the flux and variance values were taken at the pixel that corresponds to the sky
coordinates R.A.\,=\,266.4168, decl.\,=\,--29.0078 (J2000) in each sky map; 
the position of \sgra.

\subsection{\xray spectrum}
\label{s:xraySpectrum}
The \xray spectrum in the range 1--10\un{keV} was extracted using 2004 \xmm 
data from a circular region centered on the position of \igr and with
a radius of 8 arcminutes.
This radius was chosen to match the 13\am \psf full width half max (\fwhm)
of the \ibis/\isgri instrument derived from the quadratric sum of the projected pixel 
(5\am) and mask element (12\am) sizes \citep{c:gros03}. 

For a spectral extraction in which a large portion of the camera's field of view is used 
as a collecting surface, it is most suitable to use \xmm background event files
compiled from high latitude observations to construct a background spectrum.
The procedure can be summarized as follows.
\epic \pn, \mos1 and \mos2 event files are filtered to exclude all non \xray triggers 
using the event flag and pattern, after which a good time interval (\gti) selection 
is performed to exclude periods of solar flaring activity. The \gti selection criteria are 
based on the count rates in the high-energy bands, i.e.: 10--12\un{keV} for \mos1 and \mos2, 
and 12--14\un{keV} for \pn. An interval of 100\un{s} qualifies as a \gti if it has less than
18\un{cts} for both \mos cameras and 22\un{cts} in the case of the \pn camera.
These strict selection criteria ensure that only the cleanest parts of the observation 
are used. The resulting filtered event files are used to make images in the high-energy bands
in which a uniform distribution of events is expected under the assumption that energetic 
charged particles heavily dominate the instruments' respective spectra at these energies 
and that these fall uniformly on the telescope. If no point-sources are visible in these 
images, the ratios of the average count rate in the images to that of the background 
event file are used to scale the background spectra.

\section{Results}
\label{s:results}
We now come to the results we have obtained on the morphology of the Galactic nuclear 
region and on \igr, which we tentatively 
associated with the supermassive black hole \sgra in B\'elanger et al.\ (2004), and whose 
features we investigate more thoroughly in the present paper. We use three means of
investigation to study the various characteristics of the source. The mosaic provides
the fine positioning and general shape of the emission from the source and its close 
neighbours. The individual sky maps provide the elements needed for a variability study 
from kilosecond to month time scales and the average spectrum of the source can be 
used to constrain the nature of the emission.
Section\,\ref{s:mosaic} begins with a presentation of the results obtained from the 
mosaic in the range 20--40\un{keV} on the morphology of the emission from 
the central degrees.
This is followed by a discussion of the changes in the emission's morphology as a function 
of energy by looking at the mosaics in the different energy bands up to 85\un{keV},
and ends with our results on the electron-positron annihilation line at 511\un{keV}.
In \textsection\,\ref{s:lightcurves} we discuss the light curves and variability of the 
central source on different timescales, and in \textsection\,\ref{s:spectrum}  present 
the broad-band high-energy spectrum of the central arcmins of the Galaxy.
Preliminary results on the \gc with the \xray monitor \jemx are briefly discussed.

\subsection{Mosaics and spatial characteristics}
\label{s:mosaic}
\begin{figure*}[htb]
\epsscale{1.0}
\begin{center}
\includegraphics[scale=0.5, angle=-90]{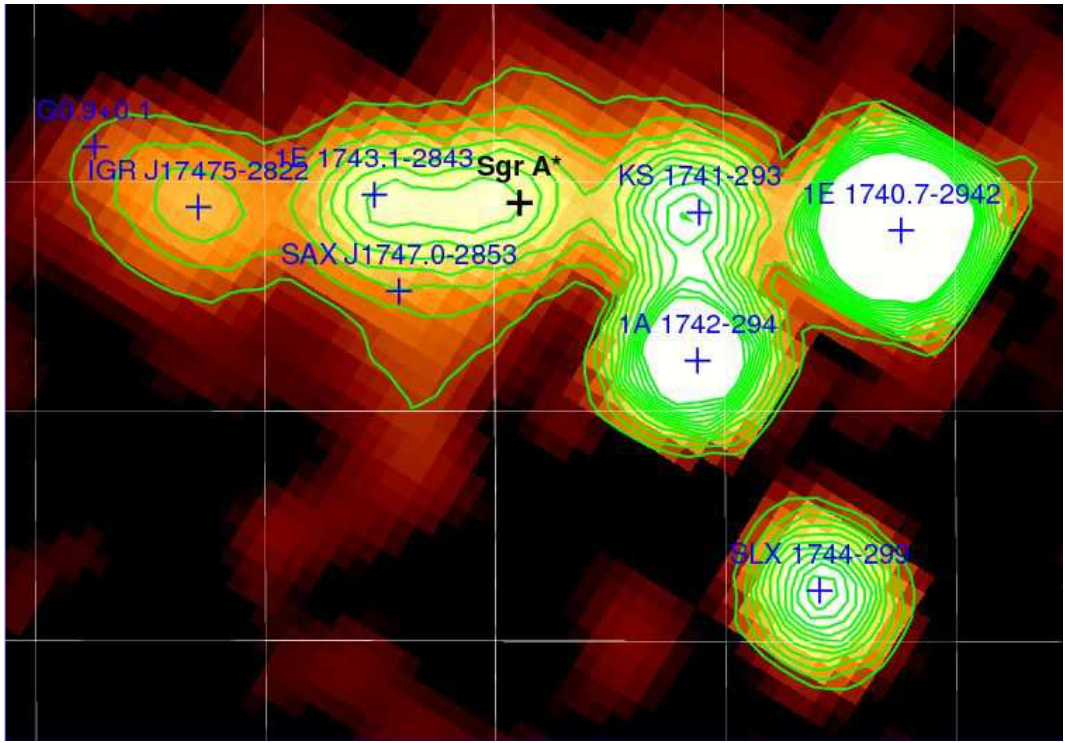}
\end{center}
\caption{\ibis/\isgri significance mosaic in the 20--40\un{keV} energy range constructed 
	from 2174 individual pointings with an effective exposure time at the position 
	of \sgra of 4.7\un{Ms}. Black indicates a statistical significance below or equal 
	to 3\,$\sigma$ and white to a significance greater or equal to 60\,$\sigma$.
	Contours mark iso-significance levels from  9.5 to 75 linearly. The
	orientation is in Galactic coordinates. The grid lines indicate galactic 
	coordinates with a spacing of 0.5 degrees.
        \label{f:mosaic}}
\end{figure*}

The mosaic shown in Figure\,\ref{f:mosaic} was constructed by summing 2174 sky images
from individual pointings and amounts to an effective exposure time of 
4.7\,$\times$\,$10^{6}$\un{s} at the position of \sgra. This Figure presents the highest 
signal-to-noise \ibis/\isgri 20--40\un{keV} image of the \gc yet published,
showing an excess of more than 45 in siginificance from  the direction of \sgra.

To model the observed morphology we have assumed that the emission is due to the
sum of the known high-energy point-sources of the region that have been detected
by \integ at least once. The main sources are 1E\,1740.7--2942, KS\,1741--293, 
1A\,1742--294,  SLX\,1744--299/300, 1E\,1743.1--2853, \igr and SAX\,J1747.0--2853.
The respective positions of these sources were derived from a simultaneous fit 
of all 8 sources in the 20--40\un{keV} mosaic. All positions were left
as free parameters except for that of SAX\,J1747.0--2853 that was fixed. This source
was quite active for the period refered to as Spring 2004 and thus contibutes to the
emission near the \gn but since its global contribution is weak and it cannot be 
clearly resolved from 1E\,1743.1--2853 we must fix its position.

\begin{figure*}
\epsscale{1.0}
\begin{center}
\includegraphics[scale=0.6, angle=-90]{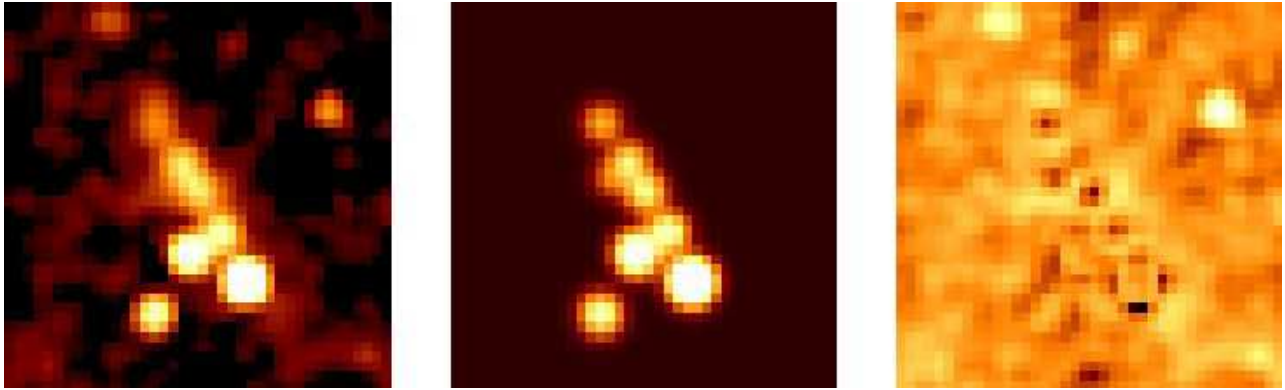}
\end{center}
\caption{Mosaic (left), model constructed from simultaneous fitting procedure
	of the 8 point-sources labeled in Figure\;\ref{f:mosaic} (center), and residuals after 
	subtraction of model from mosaic (right). These maps are oriented in Equatorial
	coordinates where North is toward the top and East toward the left.
        \label{f:fit}}
\end{figure*}

The result of the fitting procedure is 
well illustrated in Figure\,\ref{f:fit} where we see the mosaic (left), the model (middle), 
and the residual map after subtraction of the model from the sky map (right).
We can see that the spatial distribution of the  modeled image resembles very closely 
that of the mosaic, even if the residues hint at the presence of a non-uniform 
underlying emission that is not properly taken into account in the tested model.
The fitted source positions are listed in Table\,\ref{t:sourcePositions} 
where we also report the signal-to-noise, the estimated error radius corresonding to 
the 90\% confidence level and the offset with respect to the proposed counterpart.

We find that the position of \igr, detected at a signal-to-noise level of 
45 in this energy band, is 
R.A.\,=\,$\rm 17^{h}45^{m}42^{s}\hspace{-2pt}.5$, 
decl.\,=\,$-28^{\circ}59'28''$ (J2000) with an uncertainty of 0\am\bk.75.
The reliability of the derived position for this excess is supported by the fact that 
all the other sources in the field are very well positioned.
The brightest source, 1E\,1740.7--2942, is well within its associated error radius. 
For 1A\,1742--294, the reported offset is very close to the value of \psla, 
while in the case of KS\,1741--293 the best known coordinate position itself has 
an uncertainty of about 1\am \citep{c:sidoli-mereghetti99}.
The source we labeled SLX\,1744--299, is in fact a system composed of two known X-ray 
bursters located within $\sim$\,3\am of each other
\citep{c:skinner90, c:pavlinsky94, c:sakano02} and for this reason we do not 
expect the fit to yield a position within the \psla for either one of the two sources, 
independently of the detection significance. 
1E\,1743.1--2853, a well known, bright X-ray source \citep{c:porquet03a}
is almost certainly contributing to the high-energy emission in the region.
However, performing the simultaneous fit using a single source to model the
emission from the region around this source gives a centroid offset by about 2\am
from the \xmm position known to arcsecond accuracy;
a result that is not compatible with the expected error.
Since, as mentioned above, 
we know of one source detected by \isgri that was active over the course of 
the first part of 2004, namely SAX\,J1747.0--2853, we included it and fixed its position.
This yields a fitted position for 1E\,1743.1--2853 that is well
within the uncertainty derived from the source's detection significance.
The position of  IGR\,J17475--2822 was compared to the center of the Sgr\;B2 complex
and discussed below.

\igr is located at 1\am\bk.1 from the radio position of \sgra and 0\am\bk.9 from the
center of \sgraeast. It is therefore compatible with either of these sources.
Indeed, even if its associated positional uncertainty of 0\am\bk.75 is somewhat
smaller than the offset, we expect this \psla to be slightly overestimated 
when fitting multiple close sources (i.e. within the full width of the \psf). 
For example, in the case of the known source 1A\,1742--294,
the measured offset can be 20--30\% times larger than the \psla.
Morever we have found that the positions of 1E\,1743.1--2853 and \igr can change by
0\am\bk.3--0\am\bk.4 depending on the model adopted, 
and in some cases \igr is positioned only 0\am\bk.6 from \sgra.
For this reason we adopt a final error radius for the central source (and for 1E\,1743.1--2853
that has a comparable signal-to-noise ratio) of 1\am, about 30\% larger than
the \psla value of 0\am\bk.75 derived from the relation given by Gros et al.\ (2003).

On the other hand, we can safely exclude a number of other candidates
such as the transient \asca source AX\,J1746.5--2901 \cite{c:sakano02}
mentioned by Revnivtsev et al.\ (2004) as a possible counterpart for the central excess
based on the fact that it is located at a distance of more than 2\am from it.

\begin{figure*}[htb]
\epsscale{1.0}
\begin{center}
\includegraphics[scale=0.6, angle=-90]{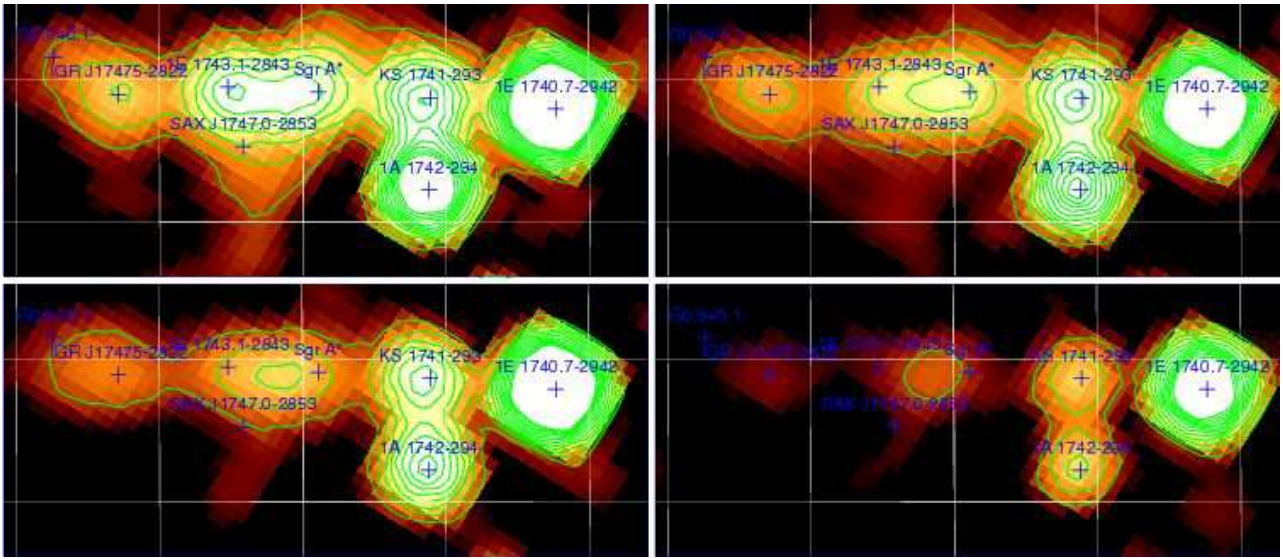}
\end{center}
\caption{\ibis/\isgri significance mosaic as in Figure\,\ref{f:mosaic} in four energy bands:
	20--30\un{keV} (top left), 30--40\un{keV} (top right), 40--56\un{keV}  (bottom left)
	and 56--85\un{keV} (bottom right). Black corresponds to statistical significance 
	below or equal to 3 and white to a significance greater or equal to 50. 
	The 12 contours levels mark iso-significance from 8 to 70 linearly.
        \label{f:mosaic_4bands}}
\end{figure*}

A similar analysis was performed on the mosaics from the same data set
in different energy bands. As is clearly seen from the iso-significance contours in 
Figure\,\ref{f:mosaic_4bands}, the morphology of the central degrees does not
radically change with increasing energy. However, we notice that the emission
that seems to bridge the sources labelled \sgra and 1E\,1743.1--2843 at low-energy
persists at higher energies such that in the 56 to 85\un{keV}
range, the emission from the region seems to be centered between the two sources.
This is a surprising result that we cannot readily interpret. An 
investigation of this based on a comparison of the emission detected by \ibis/\isgri with the 
20\un{cm} radio map, the 6.4\un{keV} Fe line-emission contours and the CS map
of the region raises several other interesting questions.

\begin{figure*}[htb]
\epsscale{1.0}
\begin{center}
\includegraphics[scale=0.6, angle=-90]{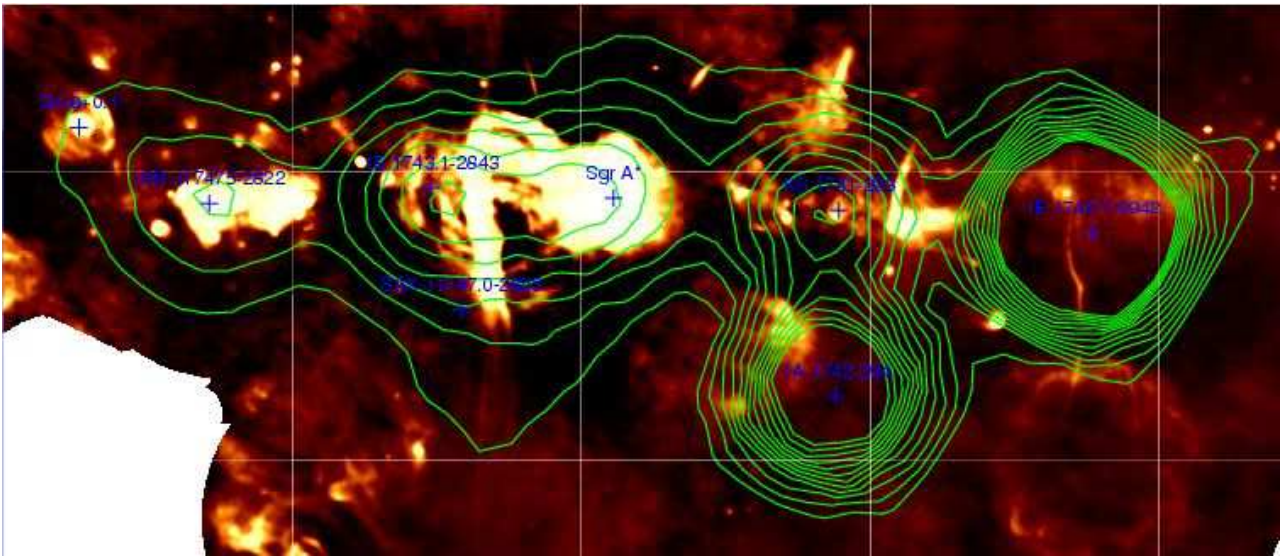}
\end{center}
\caption{Radio map of the Galactic center region at 20\un{cm}
	overlayed with the 20--30\un{keV} \isgri contours.
        \label{f:radio20cm}}
\end{figure*}

Figure\;\ref{f:radio20cm} is a radio continuum map at 20\un{cm} \citep{c:yusefZadeh04}
on which we have overlayed the 20--30\un{keV} iso-significance \isgri contours
as they are shown in Figure\,\ref{f:mosaic_4bands} (top left).
Firstly, the centroid of the very bright Sgr\,A complex which includes the luminous \sgraeast 
appears to be in best agreement with the 20--30\un{keV} \isgri contours 
(Fig.\,\ref{f:mosaic_4bands} top left). 
For completeness, we performed the simultaneous fit using the position of 
\sgraeast instead of the one for \sgra and obtained very similar results. 
The offset of the fitted source is nonetheless slightly smaller for \sgraeast that for \sgra 
but both are within the \psla and thus statistically equivalent.
We also see that the radio Arc is quite distant from the peak near \sgra and can therefore be
confidently excluded as a possible contributor to the flux at that position. However, 
the rough alignment of the radio Arc with the elongation on the 20--30\un{keV} contours in 
the direction of negative latitudes is intriguing. 
In the 56--85\un{keV} contours (Fig.\,\ref{f:mosaic_4bands} bottom right)
we see that the centroid of the emission is almost exactly between the Sgr\,A complex and the
radio Arc. This region is known to harbour large molecular clouds and there appears to be
a very good agreement between this high-energy emission feature and both 
the 6.4\un{keV} Fe line-emission, tracing irradiated molecular regions, and the CS map,
tracing regions with high gas density.
Furthermore, the centroid of this high-energy source is strikingly close to that of
the unidentified \egret source 3EG\,J1746--2851 and could in fact be its soft
\gammaray counterpart.
We extracted a spectrum for this source by fitting three sources, two of which had
their positions fixed to those of \sgra and 1E\,1743.1--2853,
and obtained a power-law photon index of $\Gamma$\,$\sim$\,2.3, and a luminosity of 
$L_{\rm X}${\footnotesize[20--120\un{keV}]}\,$\sim$\,2.6\,$\times$\,$10^{35}$\ergpersec 
for a distance of 8\un{kpc}. 
A more detailed analysis of this source will be presented in future work.

Moving in the direction of positive longitudes, we clearly see an emission region 
depicted in the 20--30\un{keV} contours and whose centroid is labeled as IGR\,J17475--2822;
a source associated with Sgr\;B2 by Revnivtsev et al.\ (2004).
Indeed this source coincides with the radio-bright Sgr\;B2 complex composed of
molecular clouds and several compact HII regions. The fitted centroid for this source,
taken to be point-like in the 20--40\un{keV} image, is positioned 1\am\bk.6 from 
the estimated center of the cloud.  As was pointed out by Revnivtsev et al.\ (2004),
there is good agreement between the 6.4\un{keV} Fe line contours and 
the emission detected by \isgri as IGR\,J17475--2822. 
Moreover, the extension toward the north could tentatively be associated with  the 
composite \snr G\,0.9+0.1 \citep{c:helfand-becker87}.
The \xray emission of its pulsar wind nebula (\pwn) has been mapped with \xmm 
\citep{c:porquet03} and the extrapolation of the flux towards 20\un{keV} of about
9\,$\times$\,$10^{-5}$\phpercmpersec, is consistent with the residual flux of roughly
0.06\un{cts/s} (20--40\un{keV}) or 8\,$\times$\,$10^{-5}$\phpercmpersec. 
It is therefore tempting to  interpret it as a detection of the highest energy
synchrotron radiation from this source that was detected for 
the first time this year by the \hess instrument \citep{c:aharonian05a}.
It is worth noting that a long \xmm exposure of this object has revealed the presence of 
variable source probably of an accreting binary type, located at a distance of 1\am
\citep{c:sidoli04}. Having a luminosity close to that of G\,0.9+0.1, 
its contribution to the residual \isgri emission could be significant.

\begin{deluxetable}{lrcccc}
\tablecolumns{6} 
\tablewidth{0pc}
\tablecaption{Positions of \integ sources in the Galactic nuclear region} 
\tablehead{ 
\colhead{Source ID} & \colhead{Signif.}& \colhead{Fitted position}& \colhead{\psla$^{\rm{a}}$} & \colhead{Offset$^{\rm{b}}$} \\
		& \colhead{}	& \colhead{(R.A., decl.)}	& \colhead{(arcmin)}	& \colhead{(arcmin)} }
\startdata

1E\,1740.7--2942 \dotfill &  241.8	&       $265.9794, -29.7430$    &  0.28 & 0.14 \\
1A\,1742--294   \dotfill	 &  98.6	&       $266.5138, -29.5109$    &  0.40 & 0.55 \\
SLX\,1744--299  \dotfill	 &  61.8	&       $266.8600,  -30.0183$   &  0.60 & 1.14 \\
KS\,1741--293  \dotfill  	 &  63.9	&       $266.2130, -29.3327$    &  0.59 & 1.23 \\
1E\,1743.1--2843\dotfill	 &  46.3	&       $266.5782,  -28.7378$   &  0.74 & 0.54 \\
\sgra	       \dotfill	  &  45.4	&       $266.4285,  -28.9918$   &  0.75 & 1.13 \\
IGR\,J17475--2822\dotfill &  18.9	&       $266.8422,  -28.4139$   &  1.24 &1.51 \\
SAX\,J1747.0--2853\dotfill & 16.5	&       $266.7500, -28.8700$    &  1.45 & 0.0 \\

\enddata 
\tablenotetext{a}{Point-source location accuracy at 90\% confidence level}
\tablenotetext{b}{Distance between the fitted and nominal source positions} 
\tablenotetext{c}{Position was fixed in the fit}
\label{t:sourcePositions}
\end{deluxetable} 

We performed the analysis of the entire data subset used to construct the mosaics in
the narrow band between 500 and 522\un{keV}. This corresponds to the \fwhm of 
the emission line in the \isgri instrument \citep{c:terrier03}. 
A background map that corresponds to this energy range was used and thus
the resulting mosaic is free of systematic effects.
No sources are detected in the field spanning 10 degrees on either side of the \gn both 
in longitude and latitude. 
We obtain a 3$\sigma$ upper limit of 1.9\,$\times$\,$10^{-4}$\phpercmpersec
to the flux from a point-source at the position of \sgra where the exposure and thus
the sensitivity is maximal. 
This limit is calculated taking into account corrections derived from the probability 
of photoelectric interaction at 511\un{keV} in the \isgri detector (34\%) and the 
fact that we have selected events using a band corresponding to 78\% of line flux. 
\spi detected a 511\un{keV} line flux of about $10^{-3}$\phpercmpersec
with intrinsic line width of 2.7\un{keV} (\fwhm) from a region well described by a Gaussian 
with a \fwhm of about 8\degree and that coincides approximately with the 
Galactic bulge \citep{c:knodlseder05}.
If we assume that our sensitivity is more or less uniform over the central 10 degrees 
around \sgra, our upper limit implies that if this emission is due to a collection of 
$n$ point-sources clustered together such that they cannot be resolved by 
\spi, then under the simplifying assumption that they all contribute 
equally to the total flux, each must have a flux of about 
$\frac{1}{n}$\;$\times$\;$10^{-3}$\phpercmpersec.
Therefore, at least 5 individual sources would be necessary to account for this 
extended 511\un{keV} emission in general agreement with the \spi result
\citep{c:knodlseder05}.

Finally, \jemx mosaics in four energy bands, 
with a total exposure is 3.2\un{Ms} and effective exposure at the 
location of \sgra of about 500\un{ks}, were constructed from 1204 science windows 
taken between 2003 February to 2004 October. 
We used the following energy ranges: 3--4, 4--8, 8--14, 14--35\un{keV} and find
no evidence in this data sample for the presence of a 
\jemx source in the Sagittarius\;A complex except for a very marginal excess in the
8--14\un{keV} mosaic.
Although this analysis is preliminary and at this point somewhat qualitative,
it is an interesting result in the light of the strong \isgri detection and obvious
intense \xray emission from this region seen by \xmm and \chandra.
It may be an additional indication that the emission is not due to a point-source
but rather to a compact diffuse emission region were thermal and non-thermal
processes take place.

\subsection{Light curves and variability study}
\label{s:lightcurves}
\begin{figure}[htb]
\epsscale{1.0}
\begin{center}
\includegraphics[scale=0.35, angle=-90]{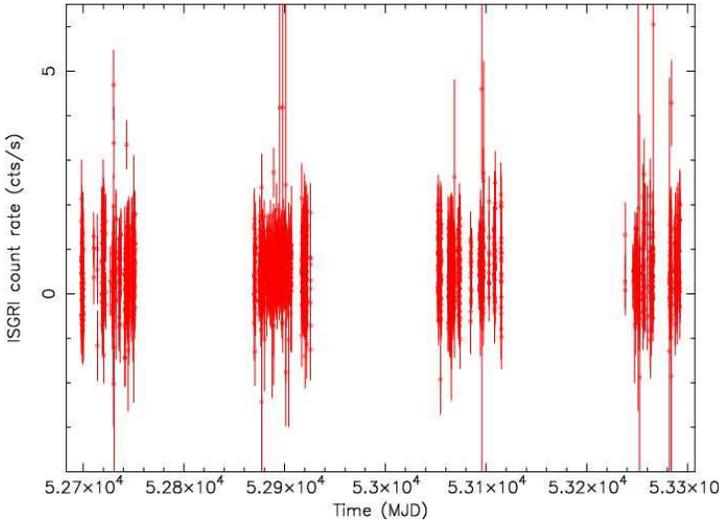}
\end{center}
\caption{Light curve of \igr in the 20--40\un{keV} band constructed from
	2174 sky images, each corresponding to one pointing (1800\un{s}).
        \label{f:unbinnedLC}}
\end{figure}

\begin{figure}[htb]
\epsscale{1.0}
\begin{center}
\includegraphics[scale=0.35, angle=-90]{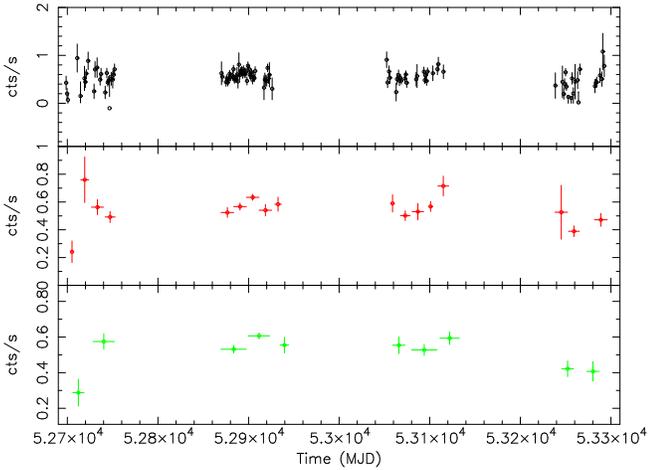}
\end{center}
\caption{Rebinned light curves of \igr constructed from the data set shown in 
	Figure\,\ref{f:unbinnedLC} with time bins of 1 day (top), 
	2 weeks (middle) and 1 month (bottom).
        \label{f:binnedLC}}
\end{figure}

The complete  light curve of \igr in the 20--40\un{keV} energy range, with a resolution of 
about 1800\un{s} corresponding to the duration of a single pointing, is shown 
in Figure\,\ref{f:unbinnedLC}. Since low amplitude variability on kilosecond time scales can 
not  be meaningfully studied for such a weak source due to statistical limitations, 
we have also done a search on longer time scales by rebinning the total light curve 
on the basis of 1 day, 2 weeks and 1 month. 
These rebinned data sets are presented in Figure\,\ref{f:binnedLC}.

No individual point deviates from the mean by more than 3$\sigma$.
The level of variability in the flux from the central source
was evaluated by means of a simple chi-squared test. 
For the unbinned data set shown in Figure\,\ref{f:unbinnedLC}, the reduced chi-squared value
is $\chi^{2}_{\rm\nu}$\,=\,1.3 (2758/2093). For the light curve with 1-day time resolution 
(Fig.\,\ref{f:binnedLC} top) we found $\chi^{2}_{\rm\nu}$\,=\,1.7 (180/109). In the case of the 2-week 
time resolution light curve (Fig.\;\ref{f:binnedLC} middle), 
the reduced $\chi^{2}$ value was found to be 3.6 (61/17). However,
if we exclude the first point in this data set which corresponds to the data collected during
revolution 46, the first observation of the \gc just after the initial calibration phase,  
we find a value of $\chi^{2}_{\rm\nu}$\,=\,2.1 (34/16), in closer agreement with the previous two.
Finally, in the case of the 1-month time resolution light curve (Fig.\,\ref{f:binnedLC} bottom) 
we find values of 5.2 (52/10) and 3.1 (28/9) if we exclude the first data point, 
heavily affected by the revolution 46 data given that there is a three-week gap between this 
revolution and the second observation of the \gc during revolution 53.
These reduced chi-squared values tend to increase as the binning gets coarser and thus
we might be seeing a small level of variability on monthly timescales.
Disregarding the data point associated with rev.\ 46, the only deviation that almost reaches
3\,$\sigma$ from the mean is at the very end of the light curve where.
This lack of evidence for significant variations in flux other than the low level 
of variability seen on monthly timescales is in contradiction with the previous detection 
of a flare from \igr \citep{c:belanger04} that we therefore do not confirm. 
We point out that those results were obtained with the preliminary analysis procedures 
and without background corrections.
The data subset covering the observation period of the reported flare (2003 April) processed 
with the most recent analysis software and background correction maps do not indicate 
significant variability with respect to the mean count rate.
Similarly, the sources 1E\,1743.1--2843 and IGR\,J17475--2822 seem rather constant
unlike the four well known \xray binaries that show very large intensity variations
over the two-year observation period.

\subsection{Spectrum of \igr}
\label{s:spectrum}
\begin{figure}[htb]
\epsscale{1.0}
\begin{center}
\includegraphics[scale=0.35, angle=-90]{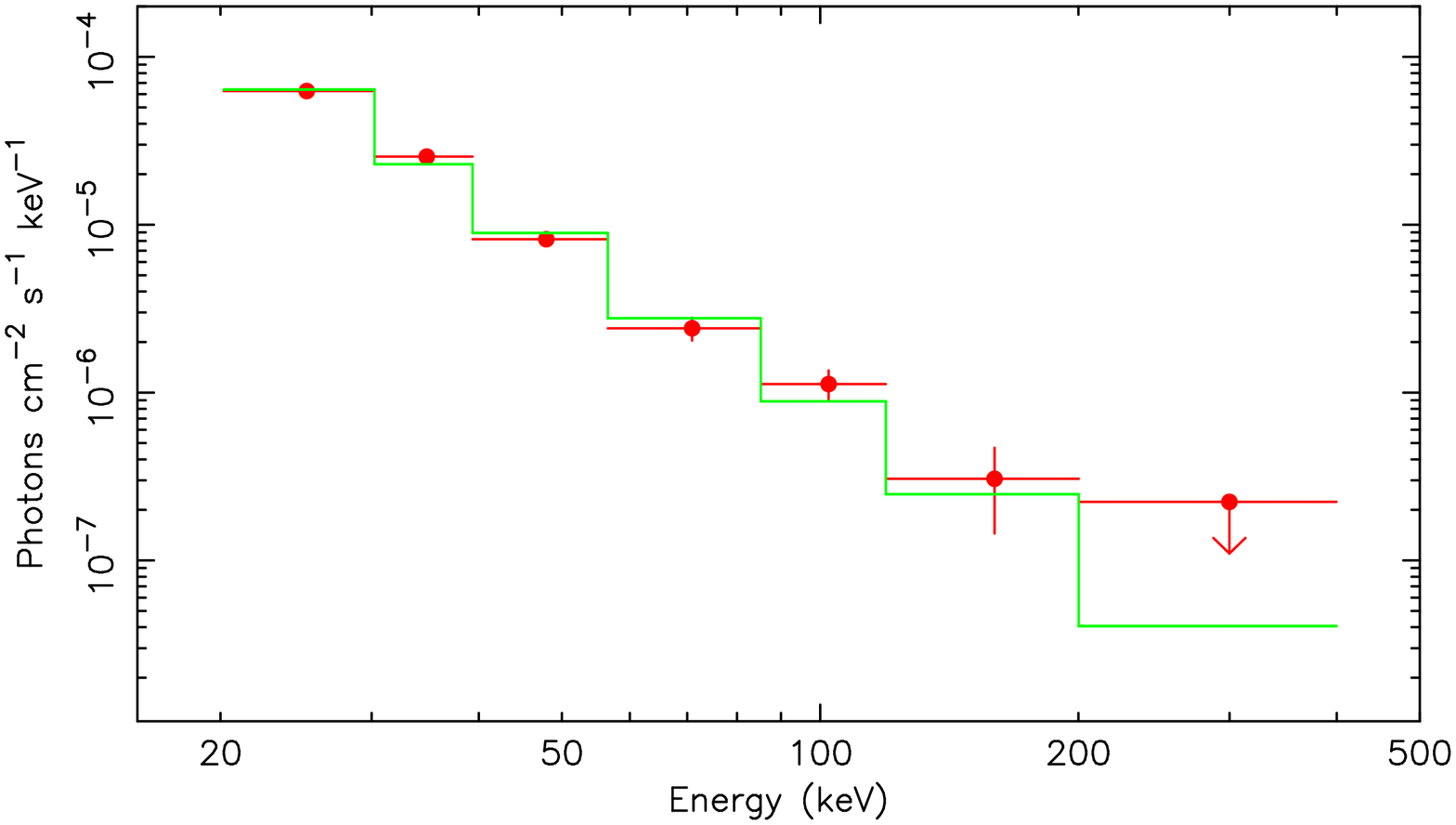}
\end{center}
\caption{Two-year averaged \isgri spectrum of the \gc source \igr in the range 
	20--400\un{kev}. The spectrum is fit with a power-law normalized over the whole
	energy range and yields a photon index of $\Gamma$\,=\,$3.04 \pm 0.08$.
	Total flux is (7.02\,$\pm$\,0.27)\,$\times$\,$10^{-11}$\ergpercmsec. 
	The last point in the spectrum corresponding to the range 200--400\un{keV}
	is given as the 1\,$\sigma$ upper limit, and the data point in the 120--200\un{keV}
	band has a significance $\sim$\,2\,$\sigma$ 
        \label{f:isgri_ufs}}
\end{figure}

Figure\,\ref{f:isgri_ufs} shows the \isgri spectrum of the \gc source that we 
modelled with a simple power-law of index $\Gamma$\,=\,3.04\,$\pm$\,0.08 
($\chi^2$\,=\,7.92 for 5 dof and 3\% systematics).  The pegged power-law 
model {\it pegpwrlw} in \xspec uses the total flux as normalization and in 
this way the photon index and normalization are independent paraters.
The total flux in the range from 20 to 400\un{keV} is
$F_{\rm X}${\footnotesize[20--400\un{keV}]}\,=\,(7.02\,$\pm$\,0.27)\,$\times$\,$10^{-11}$\ergpercmsec,
which corresponds to a luminosity of 
$L_{\rm X}${\footnotesize[20--400\un{keV}]}\,=\,(5.37\,$\pm$\,0.21)\,$\times$\,$10^{35}$\ergpersec
at a distance of 8\un{kpc} to the \gc. In the 20--100\un{keV} range, the luminosity is
$L_{\rm X}${\footnotesize[20--100\un{keV}]}\,=\,(4.56\,$\pm$\,0.10)\,$\times$\,$10^{35}$\ergpersec, 
somewhat higher than our first estimate of $\sim$\,3\,$\times$\,$10^{35}$\ergpersec 
\citep{c:belanger04}.
This is not surprising given that the first estimate was based on a rough comparison 
with the Crab's count rate in only two energy bands, 20--40 and 40--100\un{keV},
and that we now have 5 points to constrain the slope. Furthermore, 
the detection significance of the central source was much lower than in the 
present case.

\begin{figure*}[htb]
\epsscale{1.0}
\begin{center}
\includegraphics[scale=0.6, angle=-90]{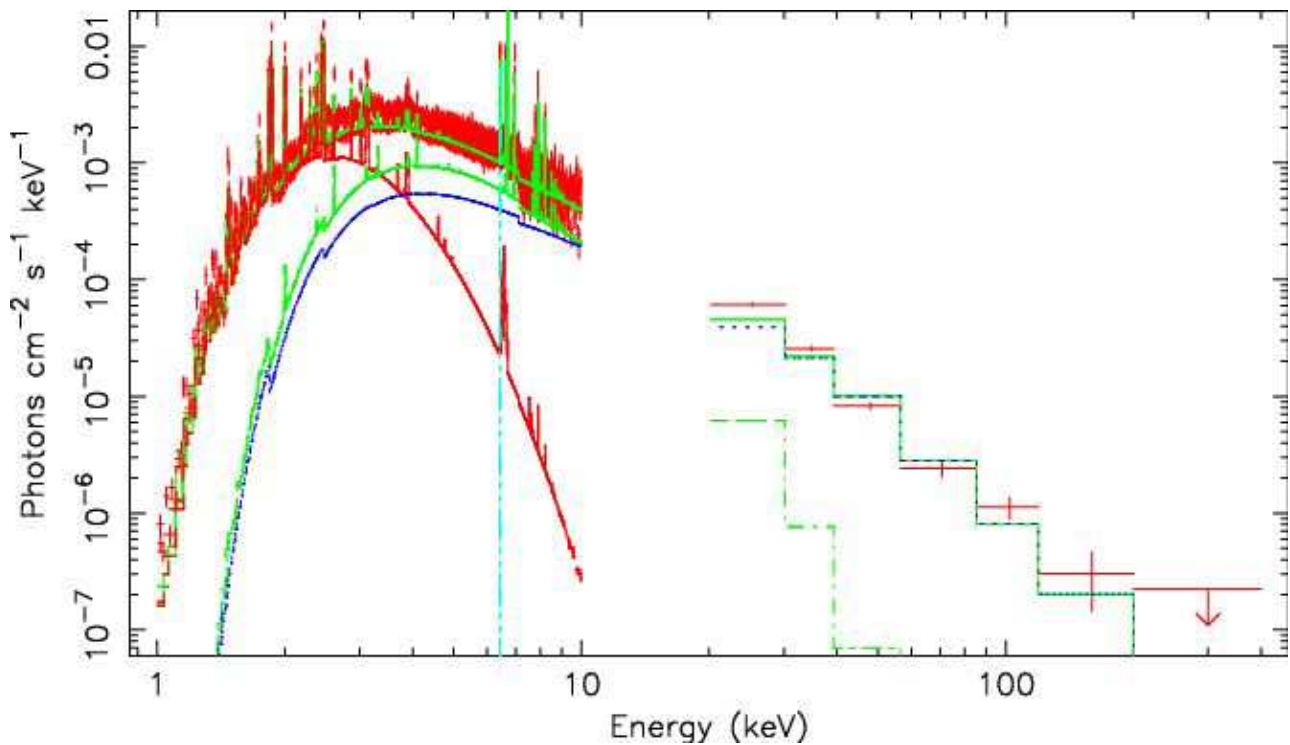} 
\end{center}
\caption{Broad band high-energy spectrum of \igr constructed from XMM-Newton data 
	in the 1--10\un{keV} energy range and from \isgri data from 20 to 400\un{keV}. 
	The \xray portion of the spectrum was build by integrating over a circular
	region of radius 8\am. 
	The model used is an absorbed two-temperature plasma plus a broken power-law
	where the low and high temperature components are drawn in red and green respectively,
	the power-law is in dark blue and the 6.4\un{keV} gaussian line is in light blue.	 
	As in Figure\,\ref{f:isgri_ufs}, the point in the last bin gives the 1\,$\sigma$ upper 
	limit on the flux in the 200-400\un{keV} range.
        \label{f:gcSpec-ufs}}
\end{figure*}

Now turning to the broad-band high-energy spectrum of \igr, 
we can see in Figure\,\ref{f:gcSpec-ufs}, the spectrum of the central source 
from 1 to 400\un{keV} where the \xray portion (1--10\un{keV}) is from
\xmm data collected during the multiwavalength campaign, and therefore 
contemporaneous with part of the \isgri data from 2004 used to construct
the soft \gammaray portion (20--400\un{keV}) of the spectrum that was discussed
earlier and is shown by itself in Figure\,\ref{f:isgri_ufs}.
The \xray spectrum was made by extracting the photon flux from a region
centered at the position of \igr and with an radius of 8\am.
This integration radius was chosen to be compatible with the \ibis/\isgri psf because
there is no obvious \xray point-source counterpart to \igr within 1\am of \sgra.
Such a point-source would have to be hard, persistant and extremely bright in \xray in 
order to be compatible with  the high-energy flux of the \integ \gc source.

The model fitting for large extended regions near the \gc is challenging for two main reasons.  
First, the \xray spectra of such extended regions give a corse, averaged spectral behaviour 
of a complex field heavily dominated by diffuse emisson that we know to have several different spectral 
components \citep{c:muno04a} but that also includes all the point sources some of which
surely contribute to the hard \xray flux.  Of course, this difficulty dissipates as
integration radius decreases since fewer components are summed together.
Second, we have no {\it a priori} knowledge of 
the nature of the emission detected as \igr and therefore do not know whether the comparison 
with the total \xray spectrum from a region that roughly corresponds to \isgri's angular 
resolution is an appropriate one.  Keeping these caveats in mind, we justify this type of 
comparison by pointing to the fact that the source coincident with the \gn, detected by \integ
as what appears to be a point-source, must undoubtedly contribute to the \xray spectrum from 
the region that corresponds to its spatial extent. The spectral transition from 
10 to 20\un{keV} must be more or less continuous and therefore we expect the high-energy 
component present in the \xray spectrum and from which \igr arises, to stand out beyond the 
thermally dominated spectrum at around 20\un{keV}. 
Therefore, a large $\chi^2$ value should not be surprising for it points to the fact
that the emission in the range from 1 to 3\un{keV} is not modelled properly for the
reason mentioned above.
Our aims in this section is to constrain the high-energy characteristics of this source
in the range 1--400\un{keV} using the \isgri spectrum above 20\un{keV}.


The broad-band spectrum can be modelled using a simple broken power-law
over the entire range to get an idea of the change of spectral index with increasing
energy. However, in order to be as constraining as possible without overlooking
possibly important components to this emssion like the hot temperature plasma
present in the \gc region, we performed the fit with
the same model as the one used by Muno et al.\ (2004a) to fit the diffuse 
emission from the various regions in the $17^{\prime}$\;$\times$\;$17^{\prime}$ 
field around \sgra referred to as Southeast, Southwest, Northwest, East, Close, and 
Northeast by the authors. This model comprises a two-temperature plasma with different 
absorption columns, a power-law and a gaussian line to fit the 6.4\un{keV} neutral Fe
emission line absorbed with the same column density. Although providing a reasonable 
fit to the data from 2--8\un{keV}, the model does not work well in the \isgri range of the spectrum. 
The power-law fit with index $\Gamma$\;$\approx$\;2 underestimates the flux in the 
20--40\un{keV} range and overestimates it above 85\un{keV}. A somewhat
better fit is provided by replacing the simple power-law with either a cutoff or broken power-law.
In the case of the cutoff power-law, we found a photon index
$\Gamma$\;$\approx$\;1 and cutoff energy of about 25\un{keV}, and in the case of the
broken power-law, the photon indeces were found to be $\Gamma_{1}$\;$\approx$\;1.5 and 
$\Gamma_{2}$\;$\approx$\;3.2 with a break energy of about 27\un{keV}.
The best fit parameters values are given in Table\,\ref{t:fitParams} where
only the free parameters are listed; all abundances are fixed to solar abundance. 


Looking closely at the unfolded spectra shown in Figure\,\ref{f:gcSpec-ufs}, 
we can distinguish the low and high temperature plasma components drawn in red
and green respectively, the gaussian line in light blue and the broken power-law
in dark blue. The hot thermal component clearly dominates the spectrum at low-energies
but its contribution is already well below that of the power-law component in 
the 20--30\un{keV} band and is totally negligible beyond that.
If we fix the temperature of the hot component at 8\un{keV},
the effect on the other parameters is small. The photon index in the cutoff power-law 
decreases from 1.09 to 1.05 and the high-energy cutoff from 24.4 to 27.7\un{keV}.
The $\chi^{2}$ value increases to 4849.0 for one more dof and therefore the reduced chi-squared
is slightly larger i.e. $\chi_{\nu}^{2}$\,=\,1.82, and the contribution of this component to the
overall flux in the 20--30 and 30--40\un{keV} bands increases by a factor of 2 but still lies 3 times
below the power-law in the first band and a factor of 7 below in the second.

\begin{deluxetable}{lcc}
\tabletypesize{\scriptsize}
\tablecolumns{7}
\tablewidth{0pc}
\tablecaption{Spectral Model of the Galactic Center Source \igr \label{t:fitParams} } 
\tablehead{
\colhead{Quantity} 			& \colhead{Cutoff PL}	     & \colhead{Broken PL}
}
\startdata
$N_{\rm H,1}$ ($10^{22}$\,$cm^{-2}$)	\dotfill 	&  $ 7.81_{-0.04}^{+0.02}$ 	& $7.79_{-0.13}^{+0.11}$\\
$kT_{\rm 1}$ (\un{keV})		\dotfill	&  $ 1.002_{-0.004}^{+0.008}$ 	& $0.99_{-0.02}^{+0.02}$\\
$N_{kT_2}$			\dotfill	&  $ 0.378_{-0.006}^{+0.009} $ & $0.38_{-0.02}^{+0.02}$\\
$N_{\rm H,2}$ ($10^{22}$\,$cm^{-2}$)	\dotfill	&  $ 13.13_{-0.23}^{+0.17}$	& $13.52_{-0.50}^{+0.56}$\\
$kT_{\rm 2}$ (\un{keV})		\dotfill	&  $ 6.56_{-0.09}^{+0.07}$ 	& $6.60_{-0.12}^{0.13}$\\
$N_{kT_2}$ ($10^{-2}$)		\dotfill	&  $ 6.35_{-0.06}^{+0.05}$ 	& $ 6.45_{-0.02}^{+0.01}$ \\
$\Gamma_{1}$	           		\dotfill	&  $ 1.09_{-0.05}^{+0.03}$ 	& $1.51_{-0.09}^{+0.06}$\\
$E_{\rm cutoff/break}$ (\un{keV})	\dotfill	&  $ 24.38_{-0.76}^{+0.55}$ 	& $27.13_{-4.39}^{2.79}$\\
$\Gamma_{2}$	           		\dotfill	&  	\nodata		& $3.22_{-0.30}^{+0.34}$\\
$N_{\Gamma}$ ($10^{-3}$\phcmskev)	\dotfill	&  $ 4.46_{-0.27}^{+0.29}$ 	& $ 7.445_{-0.001}^{+0.001}$\\
$\chi^{2}$\,(dof)			\dotfill	&  4490.7\,(2658)		        & 4458.0\,(2657) \\	
\enddata
\tablenotetext{}{Uncertainties on the parameters correspond to the 90\% confidence level}
\end{deluxetable}

\subsection{Multiwavelength campaign}
\label{s:multiwavelength}
\begin{figure}[htb]
\epsscale{1.0}
\begin{center}
\includegraphics[scale=0.35, angle=-90]{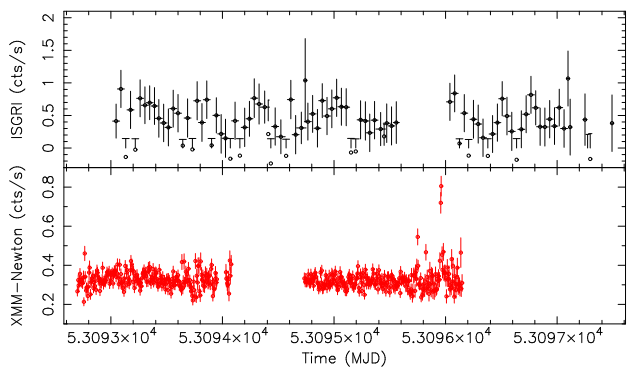}
\end{center}
\caption{\xmm light curve of the region within 10\as of \sgra in the 2--10\un{keV} 
	range with a time resolution of 500\un{s} (bottom) and \isgri light curve
	of \igr in the range 20--30\un{keV} (top) with time resolution of about 
	1800\un{s} for epoch\,1 of the multiwavelength campaign. The data gaps
	correspond to the time between orbits for which there is no scientific data.
        \label{f:xmm-integ1}}
\end{figure}

\begin{figure}[htb]
\epsscale{1.0}
\begin{center}
\includegraphics[scale=0.35, angle=-90]{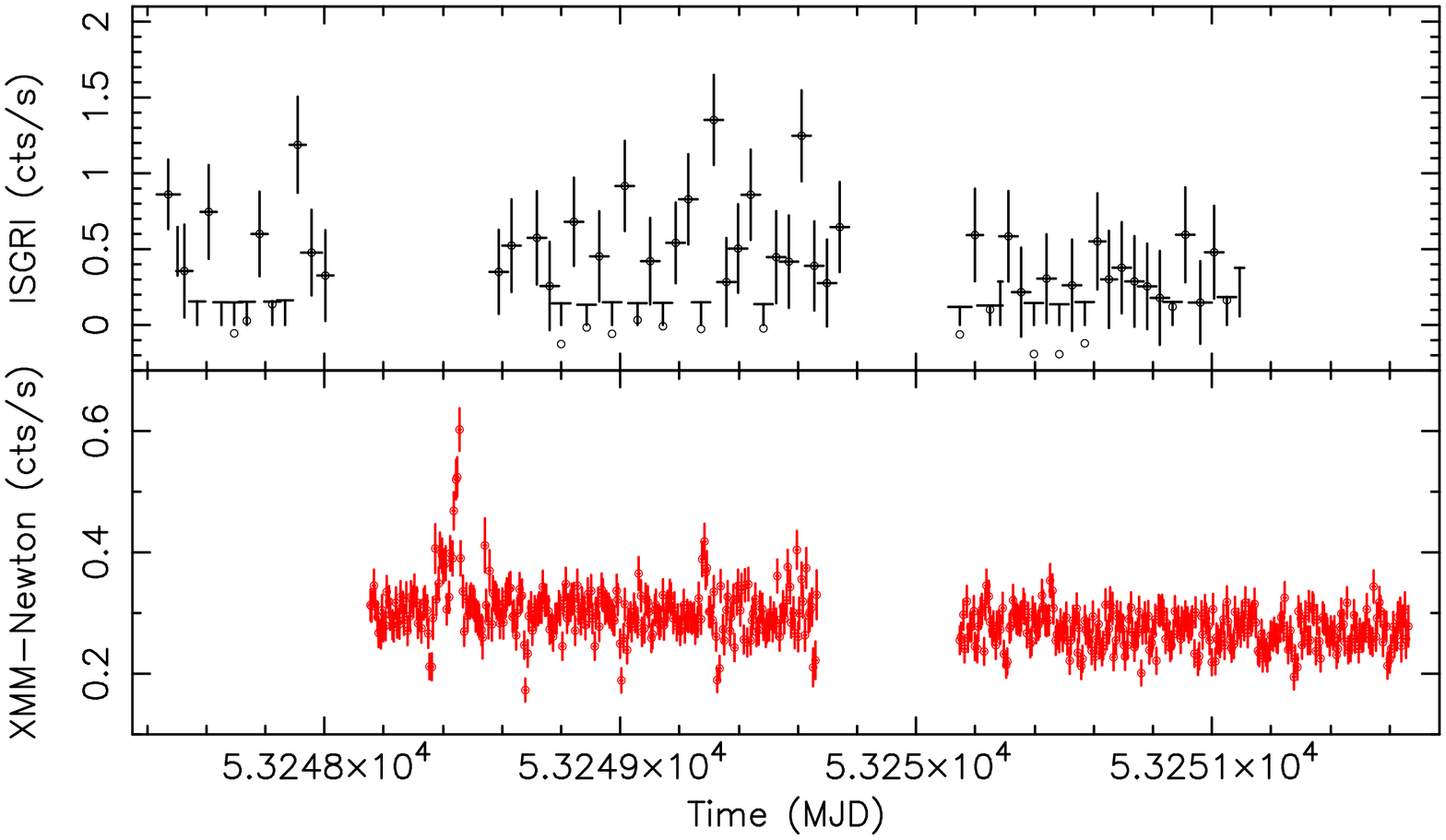}
\end{center}
\caption{Same as in Figure\,\ref{f:xmm-integ1} for observation epoch\,2. The
	clear periodic dips in the \xmm light curve are caused by the
	eclipses of the transient \cxo \citep{c:porquet05b, c:muno05b}.
        \label{f:xmm-integ2}}
\end{figure}

A multi-wavelength campaign to study \sgra with a total exposure time of about 500\un{ks}
was performed in two segments, the first of which was from 2004 March 28 to April 1
and that we refer to as epoch\,1, and the second from 2004 August 31 to September 3
refered to as epoch\,2. The primary aim of this campaign was to study 
correlated variability, particularly in the IR, \xray and soft \gammaray energy bands.
Figures\,\ref{f:xmm-integ1} and \ref{f:xmm-integ2} show the 2--10\un{keV} \xmm light curve of 
a 10\as region around \sgra in the bottom panels, and the 20--30\un{keV} \isgri light curve of \igr in 
the upper panels for epochs\,1 and 2 respectively. As is clearly visible, the periods during which 
the factor-40 flares from the direction of \sgra occured do not have simultaneous coverage 
in the \integ data. Unfortunately, both data gaps in the \isgri light curve correspond to the period 
between orbits. For this reason we are still unable to conclude whether or not we can expect
to detect a correlated variability in the \xray and soft \gammaray bands. There are two 
features worth mentioning that can be noticed in Figure\,\ref{f:xmm-integ2} although they
have marginal statistical significance. First, two points in the \isgri light curve,
approximately in the middle of the upper panel, stand out at about 2.5\,$\sigma$ above the mean.
These are temporally coincident with the two hiccups at the end of the first data subset in 
the \xray light curve shown in the bottom panel. 
Second, the weighted mean \xray count rate is somewhat higher in the first 
data subset ($0.30\pm 0.001\un{cts/s}$) than in the second ($0.27\pm 0.001\un{cts/s}$), 
a behaviour apparently seen also in the two corresponding segments of the \isgri light curve 
where the weighted mean count rate is $0.42\pm 0.03\un{cts/s}$ in the first 
and $0.20\pm 0.07\un{cts/s}$ in the second. 
This indicates that there may be a relationship between the flaring activity of \sgra and
emission at higher energies. Future simultaneous observations will undoubtedly help
elucidate this point which remains uncertain.

\section{Summary and Discussion}
\label{s:discussion}

\subsection{Summary}
We have studied the morphology of the high-energy emission from the central few degrees
of the Galaxy in the energy range from 20 to 400\un{keV} based on a sample of
\integ data collected from 2003 February to 2004 October with a total livetime of
7\,$\times$\,$10^{6}$\un{s}. We paid particular attention to the characteristics of the 
emission from the Galactic nuclear region where we detect a source with high
significance in the 20--40\un{keV} energy range located at 
R.A.\,=\,$\rm 17^{h}45^{m}42^{s}\hspace{-2pt}.5$, 
decl.\,=\,$-28^{\circ}59'28''$ (J2000) with an uncertainty of 1\am and therefore
compatible with the position of the central black hole \sgra. This detection
confirms the results obtained by B\'elanger et al.\ (2004) on the \gc source \igr.

The source \igr is persistent and shows no variability at the 3$\sigma$ level in
contradiction with what was suggested to in B\'elanger et al.\ (2004). This 
result holds at kilosecond, daily, bi-weekly and monthly timescales.

The spectrum of the central source in the 20--400\un{keV} range is well fit by
a power-law of index $\Gamma$\;$\approx$\;3. We have combined this dataset 
with the \xray spectrum of a circular region with radius 8\am centered the \integ \gc 
source derived from partially contemporaneous \xmm data collected during
observations of the \gc performed in 2004 in the range 1--10\un{keV}.
From this we find that the broad-band high-energy spectrum can be fit equally well
with a model that comprises a two-temperature plasma 
($kT_{1}$\;$\approx$\;1.0, $kT_{2}$\;$\approx$\;6.5), 
a Gaussian line to account for the neutral Fe emission at 6.4\un{keV}, 
and either a cutoff power-law with photon index $\Gamma$\;$\approx$\;1 and 
cutoff energy of about 25\un{keV} or a broken power-law with photon indeces
$\Gamma_{1}$\;$\approx$\;1.5 and $\Gamma_{2}$\;$\approx$\;3.2, and 
break energy $\approx$\,27\un{keV}.

We also detect hard ($\Gamma$\,$\sim$\,2.2) emission from a region located 
between \sgra and the radio Arc that seems to coincides with the 6.4\un{keV}
emission from neutral to weakly ionised Fe and with the CS map of the region. 
As is the case with IGR\,17475--2822,
we believe that this new detection of hard \xray emission orginates in one
or several large molecular clouds known to exist in that region.

The nature of the emission from the direction of the Sgr\,A complex detected 
as \igr is unknown. In what follows we discuss a number of scenarios in an attempt
to identify the source of the emission detected by \integ and apparently 
centered on \sgra.

\subsection{Hot plasma}
\label{s:plasma}
The hot component at 6--8\un{keV} of the two-temperature plasma at the \gc 
is well known and its presence is viewed by many as problematic in terms of it
being confined given the escape velocity of a Hydrogen plasma at that temperature 
or of the heating mechanisms that would be required to supply energy to the plasma
were it not confined. It is interesting to ask what is this hot plasma's extrapolated
flux at energies between 20--30\un{keV} in order to assess its possible contribution
to the flux detected from the central arcminutes. In \textsection\,\ref{s:spectrum}
we showed that although the hot plasma component totally dominates the spectrum 
in the range from 2 to 5\un{keV}, its contribution to the overall flux is several times less
than that of the power-law in the 20--30\un{keV} range and drops to a negligible level
with respect to the power-law component in the range 30--40\un{keV}. 
These estimates are based on the best fit value for the temperature of 6.5\un{keV}
and clearly the contribution increases somewhat if we fix the temperature at 8\un{keV}.
Nonetheless this hot plasma can only provide a very small fraction of the emission 
detected from the direction of the \gn by \integ, and this solely in the first band,
from 20 to 30\un{keV}.

\subsection{\xray transient sources}
\label{s:transients}
A large number of \xray transients have been detected in the neighbourhood of \sgra
in the last few years \citep{c:muno04b, c:porquet05a}. In fact, there even appears to be 
an over-abundance of such sources near the \gc where four \xray binaries lie within 
1\un{pc} (25\as) of \sgra \citep{c:muno05a}. These sources, observed at luminosities between
$10^{33}$ and $10^{36}$ \ergpersec in the 2--10\un{keV} range, are not particularly 
bright since these luminosities are intermediate between quiescence and outburst. 
In the paragraphs that follow we attempt to estimate the effect of such sources on 
the flux of \igr in the 20--120\un{keV} range.
We restrict ourselves to this energy range because it is here that we have the
best estimate of flux from the \isgri data.

The \gc source \igr shows very little variability and hence its persistant quality
demonstrates that it is probably not heavily influenced by the sometimes radical brightening
of an individual \xray transient as it moves from quiesence to outburst.
Furthermore, the high detection significance in the 20--40\un{keV} band permits
an accurate determination of the emission centroid and yields an uncertainty of
about 1\am. For this reason, we do not consider transients that lie outside the
error radius of \igr but turn our investigation to the 4 transients detected by \chandra
within 30\as of the central \bh \citep{c:muno05a} and pay particular attention to
the remarkable \lmxb located just 3\as south of \sgra \citep{c:muno05b}.
This transient, \cxo, was very bright during both epochs of the multiwavelength campaign,
the only period for which we have contemporaneous \xray and \integ data, and will
therefore be the main focus of the discussion.

\begin{figure}[htb]
\epsscale{1.0}
\begin{center}
\includegraphics[scale=0.35, angle=-90]{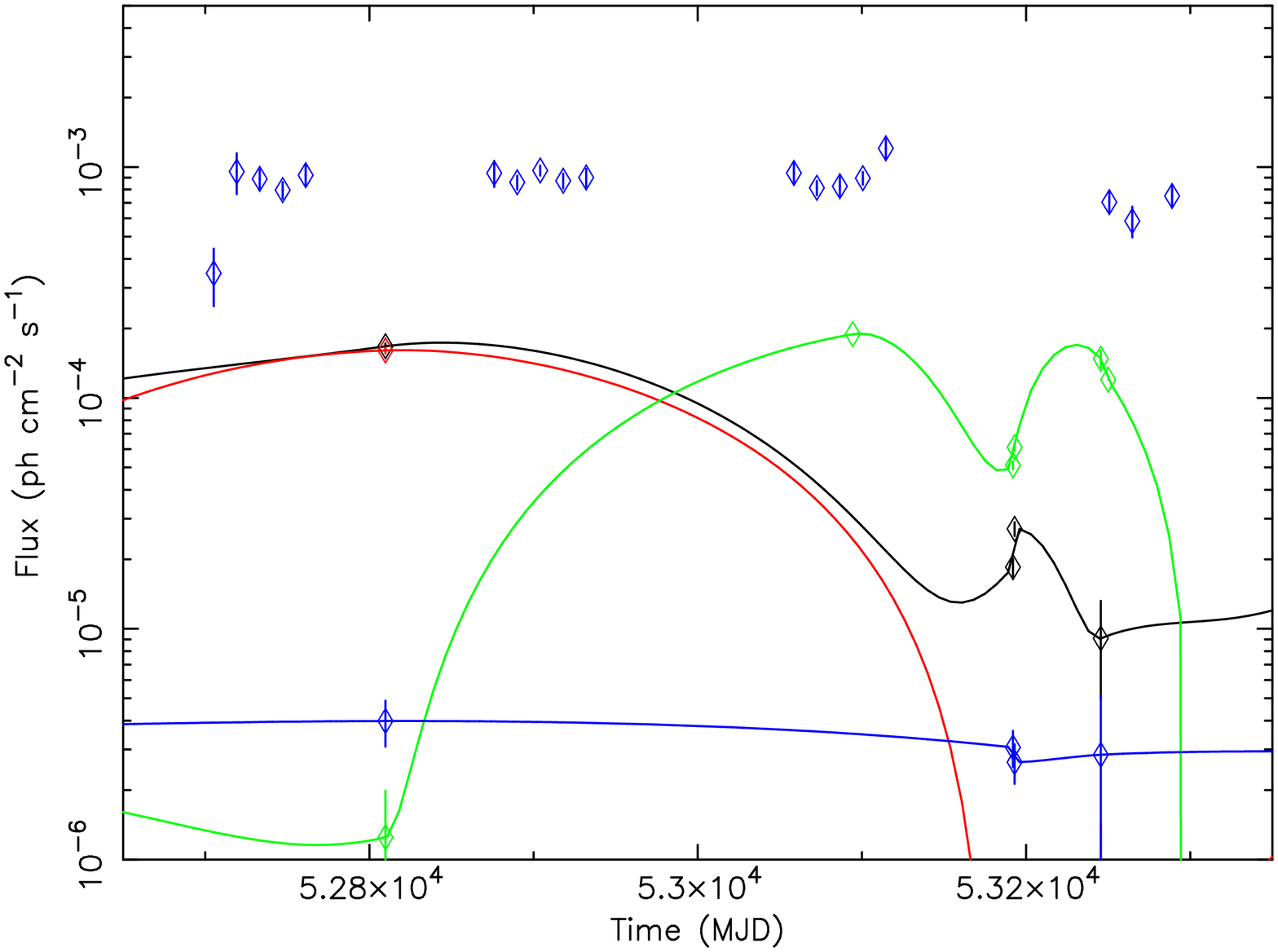}
\end{center}
\caption{Light curves in the 2--8\un{keV} energy range of \cxogc174535.5-290124 (black),
	\cxogc174540.0-290005 (red), \cxo (green), \cxogc174538.0-290022 (blue)
	and in the 20--40\un{keV} range for \igr binnned in 2-week periods. 
        \label{f:transients}}
\end{figure}

The four transients detected within 30\as of \sgra are also the hardest of the
seven transients discussed by Muno et al.\ (2005a). These are:
\cxogc174535.5--290124, \cxogc174538.0--290022,
\cxogc174540.0--290005 and \cxogc174540.0-290031.
Using the available \chandra data for these (and \xmm for \cxo),
we constructed their respective light curves in the range 2--8\un{keV} 
and rebinned to make the long term trends more apparent 
(the \chandra results were published in Muno et al.\ (2005a)).
We plotted the light curve for \igr rebinned in 2 week segments on the same time line. 
The 20--40\un{keV} range is used because it has the highest signal-to-noise ratio. 
Figure\,\ref{f:transients} shows these light curves where that of \igr appears as the 
top most. The flux is in units of \phpercmpersec and the time is in modified Julian days. 

The observation cycles are clearly depicted by the \isgri light curve where each
data subset corresponds to a season, the first one being Spring 2003 and the 
last, Fall 2004. As was found in \textsection\,\ref{s:lightcurves}, there is no clear
sign of variability in the \integ light curve other than the close to 3$\sigma$ deviation
in the very last part of the light curve. Furthermore, it is difficult to draw conclusions
from the general comparison with the flux levels of the four hardest \chandra transients 
shown in Figure\,\ref{f:transients}. 
We will now take a closer look at the unusual transient \cxo.

This source was discovered during the 2004 July \chandra observations of the \gc 
\citep{c:muno05b} and was also seen to be active throughout both epochs of the 
multiwavelenth campaign of 2004, reaching its peak \xray brightness in April. 
A very clear periodic signal was detected from this source during epoch\;2 and the
orbital period was determined to be about 8\un{h} \citep{c:porquet05b, c:muno05a, c:belanger05}.
This source, located remarkably close to the supermassive \bh at only $\sim$\,3\as 
from the radio position of \sgra, was observed in radio during the campaign \citep{c:bower05}, 
and seen to produce a jet with luminosity $L_{\rm jet}$\,$\sim$\,$10^{37}$\ergpersec. 

In order to estimate the possible contribution of this source to the \integ flux,
we used the spectral parameters derived from the contemporaneous \xmm observations
and presented in Porquet et al.\ 2005. Taking a photon index of $\Gamma$\,=\,1.6
and 2--10\un{keV} flux normalization of 1.3\,$\times$\,$10^{-12}$\ergpercmsec
we find that the extrapolated 20--120\un{keV} luminosity is about 5\,$\times$\,$10^{34}$\ergpersec. 
For an index $\Gamma$\,=\,1.98 and flux $F_{\rm X}$\,=\,1.8\,$\times$\,$10^{-12}$\ergpercmsec,
we get an extrapolated luminosity of 2.5\,$\times$\,$10^{34}$\ergpersec.
The largest of these values amounts to about one tenth of the derived luminosity of \igr 
and is probably still over-estimated given that the spectrum of a \lmxb in outburst is rarely 
a pure power-law above energies $\sim$\,100\un{keV}. 
A more realistic description would be a comptonization model where 
the seed photons are boosted to higher energies as they encounter fast thermal electrons 
in the hot (30--50\un{keV}) corona. In this case, the contribution in the \isgri band would 
very likely be even less than the above extrapolations. 
According to this simple estimate, ten sources similar to \cxo in luminosity and spectral index
within a few arcminutes of \sgra would have 
to be active more or less continuously to account for the 20--120 luminosity of \igr.
Of course, the flux could be due to a substantially larger number of dim and hard sources 
that have until now evaded detection by \chandra and \xmm and that are clustered around 
\sgra. This however, is an unlikely scenario.

We have considered and tried to quantify the effect of the known \xray transients
located within 30\as of \sgra on the \isgri count rate. We find that the flux
of \igr can not be explained by the contribution of transients located very close 
to the central \bh, and that their contribution appears to be negligeable over the range
from 20 to 120\un{keV}.

\subsection{Flares from \sgra}
\label{s:flares}
Our campaign to search for correlated variability in the \xray and soft \gammaray flux
from the central \bh was inconclusive due to \integ data gaps during the \sgra flares.
Of the two factor-40 flares that occured on 2004 March 31 and August 31,
the first was somewhat harder with a photon index of $\Gamma$\,=\,1.52
compared to $\Gamma$\,=\,1.87 for the second flare \citep{c:belanger05}.
By extrapolating the flux of the hardest flare to the 20--30\un{keV} energy band
we find that its contribution to the \isgri count rate using an average 
effective area in this band of 650\un{cm^2} should be about 0.05\un{cts/s}.
Taking a peak flare flux equivalent to twice the average gives 0.1\un{cts/s} 
and so since \igr's observed \isgri count rate in this range is around 0.4\un{cts/s}
we would expect a rise of $\sim$\,25\% due to the flare.
Based on these simple assumptions, 
this type of event would therefore not be detectable by \isgri, and for the same
reason, cannot explain the emission seen as \igr.

Furthermore, the derived luminosity of the \gc source of around 5\,$\times$\,$10^{35}$\ergpersec,
cannot result directly from the integration of successive flares from the central black hole. 
The flares occur on average once per day and have typical luminosities of $10^{35}$\ergpersec for 
durations of a few thousand seconds. If the flares last 3 to 20\un{ks}, even if they all reached
peak luminosities of around $10^{36}$\ergpersec, this would still not be enough to make 
up the persistant luminosity of \igr. This is not to say, that the acceleration of particles 
to very high lorentz factors during such a flare could not lead to a secondary high-energy
emission that would not be detected by \xray instruments but that would contribute 
to the flux in the range 20--400\un{keV}.

\subsection{Charged particle acceleration} 
The detection of a persistant source up to about 120\un{keV} compatible with the position
of the central \bh raises the very interesting possibility that it may be related to the
TeV source detected from the same region by \hess \citep{c:aharonian04}. 
These observations lend crucial support to the idea that acceleration of particles to very high
energies is taking place at the \gc \citep{c:crocker05}. Furthermore, all of them
agree on the apparent absence of variability from the central source.

The \hess Collaboration has been particularly successful at determining the high-energy
properties of this source detected over two epochs, 2003 June-July and 2003 July-August,
and together with the earlier 30\un{MeV} to 10\un{GeV} \egret detection of
the continuum source 3EG\,J1746--2851 within 1\degree of the nucleus \citep{c:mayer98},
this TeV detection provides some evidence of hadronic acceleration at the \gc,
either by \sgra itself, within the shocked shell of an \snr, such as \sgraeast or by some 
other mechanism inlcuding the interaction of non-thermal filaments with dense molecular 
environments. Protons could be accelerated, either via $1^{\rm st}$ order Fermi
acceleration at a shock, or via stochastic acceleration (a $2^{\rm  nd}$ order Fermi process)
in a turbulent magnetic field, and then scatter with ambient protons to produce pions. 
The $\pi^0$'s subsequently decay into 2\,\gammarays, whereas the $\pi^+$'s and 
$\pi^-$'s initiate a muon, electron, positron, and neutrino cascade 
(see e.g., Markoff, Melia \& Sarcevic 1997,1999). Some evidence for a pion origin of 
the \gammarays is provided directly by the \egret spectrum, which exhibits a clear break 
at $\sim$\,1\un{GeV}, and therefore cannot be fit by a single power law. 
Instead, this break is usually attributed to the rest-frame energy of pion-decay photons.
The secondary electrons and positrons produced by the charged pions in concert with the
$\pi^0$-decay photons, are capable of producing their own \gammaray emission via
bremsstrahlung and Compton scattering.  
For example, if these leptons build up to a steady-state distribution balanced by 
bremsstrahlung and Coulomb losses, they naturally account for the lowest energy 
\egret datum.  This is rather important because the pion decays link the lepton 
and photon generation rates, so the bremsstrahlung and
pion-decay photon emissivities are tightly correlated.

The possibility that the relativistic protons may be accelerated close to \sgra was
first explored by Markoff, Melia \& Sarcevic (1997, 1999), who concluded that the
maximum attainable energy is of the order of 4\,$\times$\,$10^{5}$\un{TeV}.
However, there appear to be two principal reasons why the $pp$ scatterings that lead 
to pionic \gammaray emission probably do not occur in the acceleration zone itself. 
First, the ensuing particle cascade would produce a copious supply of energetic leptons 
that, in the presence of the inferred $\sim$\,1--10\un{G} magnetic field for this source,
would greatly exceed \sgra's observed radio flux. 
Second, the lack of variability in the data from keV to TeV energies, 
argues against a compact point-source like \sgra. The more recent analysis
of proton acceleration within \sgra (Liu, Petrosian \& Melia 2004; Liu, Melia
\& Petrosian 2005) has shown that these relativistic particles actually diffuse to
distances $\sim$\,2--3\un{pc} away from the acceleration site before scattering with the
ambient protons, and therefore the ensuing leptonic cascade does not overproduce
emission at longer wavelengths having left the region where the magnetic fields are very strong. 
In support of this picture, wherein the relativistic particles responsible for the \hess and 
\integ source are accelerated near \sgra, it is worth noting that the mechanism responsible
for accelerating the electrons required to account for \sgra's 7\un{mm} emissivity, 
also accelerates protons in the system. These protons do not radiate as efficiently 
as electrons and therefore diffuse away from the acceleration site.  
The electron acceleration rate implied by the radio measurements also corresponds to the
right $\sim$\;TeV luminosity from the protons flooding the region surrounding the black hole 
to match the \hess measurements. In addition, the time required for these particles to diffuse
outwards is $\sim$\,$10^5$--$10^6$ years, which would argue against any rapid variability
in the TeV \gammaray emission, as observed. Thus, although both the \hess and \integ
\gc sources appear to be slightly extended, the origin of the particles responsible for 
the broad band spectrum of the nuclear region may ultimately still be \sgra.

\subsection{\sgraeast}
The supernova renmant \sgraeast whose centroid is located 50\as from \sgra, 
is another likely source of \gammarays near the \gn (see Melia \& Falcke 2001).
It is a member of a class of remnants detected at 1720\un{MHz} 
(the transition of OH maser emission), a signature of shocks produced at the interface between
supersonic outflow and the dense molecular cloud environments with which they interact.
It has already been shown \citep{c:fatuzzo-melia03} that \sgraeast is capable of producing
the observed \gammaray luminosity detected by \egret once the unusually-high ambient particle
density ($>$\,$10^{3}$\un{cm^{-3}}) and strong magnetic field ($>$\,0.1--0.2\un{mG})
are taken into account. In a thorough examination of the particle acceleration and energetics
in this source, Crocker et al. (2005) demonstrated that \sgraeast could very well also
be the source of the TeV spectrum measured by \hess.  One should note, however, that the \egret
and \hess sources are probably not coincident. The centroid of the \egret emission
\citep{c:mayer98} appears to be significantly displaced away from \sgra, whereas
the TeV source lies within $\sim$\,1\am. In addition, although the \egret and \hess spectral
indices are similar ($\sim$\,2.2), the corresponding fluxes at GeV and TeV energies differ by
over an order of magnitude. It appears that the \egret source must cutoff well below the TeV range
suggesting, together with the relative spatial displacement of the two sources, that we must
be dealing with at least two separate regions of \gammaray emissivity.  

\sgraeast may in fact be a very good candidate as the source of the emission
detected by \integ and \hess at the Galactic center.
This composite \snr in its radiative phase has several observational characteristics akin to
supernovae in molecular clouds or dense environments: a bright radio shell with a strong 
non-thermal synchrotron component, \xray emission from the compact central core dominated 
by the hot thermal plasma component and a strong He-like 6.7\un{keV} Fe emission line.
It shares some of these features with G\,0.9+0.1 that was recently confirmed as a source
of TeV radiation by the \hess collaboration \citep{c:aharonian05a} and its local environment
make it a good candidate for powerful particle acceleration. There is one important distinction
between G\,0.9+0.1 and \sgraeast however; the compact core of G\,0.9+0.1 is known to be a
\pwn for which is central pulsar has been detected \citep{c:gaensler01} and that has a hard,
non-thermal \xray emission. The \xray core of \sgraeast is so highly dominated by thermal 
emission that non-thermal \xray emission does not even seem to be present. We know however, 
that such behaviour is not unexpected for \snr in molecular clouds or dense environments 
\citep{c:bykov02}. Moreover, there is preliminary evidence that several of the new \hess 
sources detected during the Galactic plane scan might be \snr with weak \xray emission
\citep{c:aharonian05b}.

\subsection{Final comments}

The \isgri instrument on the \integ satellite has detected with a high significance hard
\xray and soft \gammaray emission centered within 1\am of the Galactic nucleus. 
We have analyzed two years of \integ observations and thoroughly examined this data 
over the energy range from 20 to 400\un{keV}. Combining these results with \xmm data 
in the energy range from 1 to 10\un{keV}, we have found that this emission
cannot be attributed to the hot thermal plasma in the Sagittarius complex,
it cannot be explained by the integrated flux from known \xray transients near the 
central black hole, and it cannot be the simple extrapolation of the \xray flux of 
flares from \sgra. The fact that \igr is comparable in brightness to the well known binary 
system 1E\,1743.1--2853 in the 20--40\un{keV} range but that 
unlike this source it does not produce the large \xray flux that makes
1E\,1743.1--2853 so incredibly conspicuous in the soft \xray band suggests that
\igr is not point-like. In addition, the fact that \jemx, with its $\sim$\,3\am angular
resolution, easily detects the known binaries in the
region including 1E\,1743.1--2853 but does not see any emission from the \gn
also suggests that the nature of \igr is not point-like. These considerations lead us
to conclude that \igr is a compact diffuse emission region a few arcminutes in size
and where astrophysical processes give rise to thermal and non-thermal emission.

The power of investigation and discovery at our disposal through combining the use of
very high resolution \xray instruments like \xmm and \chandra, radio observatories like the 
\vla, and the new generation of \gammaray telescopes like \integ and \hess at the highest energies
is astounding. To be faced with data from observations that cannot be readily explained nor 
understood is stimulating and inspiring. This is the kind of puzzle that we face in trying to 
understand the nature of the emission detected from the direction of the Galactic nucleus with 
\integ and \hess.  

Finally, In section\,\ref{s:mosaic} we discussed the 20--40\un{keV} morphology of the emission near the 
\gc and due to the very long effective exposure of 4.7\un{Ms}, the maps in the different 
energy bands revealed that 1) as the brighter and softer sources \igr and 
1E\,1743.1--2853 become fainter with increasing energy, a hard but dim emission about 6\am
from \sgra in the direction of positive longitudes appears, and 2) that the spectral
index of this emission is quite hard and apparently qualitatively similar to that
of  IGR\,J17475--2822 (Sgr\,B2). This new source of soft \gammarays could well be closely 
related to the unidentified \egret source 3EG\,J1746--2851 and to the giant
molecular cloud G\,0.13-0.13.

\acknowledgements{We would like to thank Anne Decourchelle and Jean-Luc Sauvageot
	for their precious help with the \xmm spectral extraction and analysis,
	Andrei Bykov and Francois Lebrun for useful and interesting discussions, and
	Micheal Muno for providing the \chandra data on the transients and discusing
	the possible contributions from these to the flux of the central \integ source, and
	Katsuji Koyama for providing us with ASCA map of the \gc region at 6.4\un{keV}.  
	G.\ B\'elanger is partly supported by the French Space Agency (CNES).}

\end{document}